\documentclass[useAMS,usenatbib]{mnras}
\usepackage[utf8]{inputenc}
\usepackage[T1]{fontenc}
\usepackage{epsfig,times,amssymb,amsmath,verbatim,xspace}
\usepackage[usenames,dvipsnames,svgnames,table]{xcolor}
\usepackage{tablefootnote}
\usepackage{todonotes}
\usepackage{tabularx}
\usepackage{hyperref}
\usepackage{float}
\usepackage{hhline}
\usepackage{lineno}
\usepackage{multirow}
\usepackage{natbib,ifthen,soul}
\usepackage{orcidlink}

\usepackage{academicons}
\newcommand{\orcid}[1]{\href{https://orcid.org/#1}{\textcolor[HTML]{A6CE39}{\aiOrcid}}}

\newcommand{\omegam}{\ensuremath{\Omega_\mathrm{m}}}
\newcommand{\omegab}{\ensuremath{\Omega_\mathrm{b}}}

\newcommand{\as}{\ensuremath{A_\mathrm{s}}}
\newcommand{\ns}{\ensuremath{n_\mathrm{s}}}
\newcommand{\lcdm}{$\Lambda$CDM}

\newcommand{\blockfont}[1]{{\textsc{#1}}\xspace}

\def\rmd{\mathrm{d}}

\title[An empirical approach to model selection: weak lensing and intrinsic alignments]{An empirical approach to model selection: weak lensing and intrinsic alignments}

\author[A.~Campos et al]{
A. Campos,$^{1,2}$\thanks{andresar@andrew.cmu.edu}\orcidlink{0000-0002-5124-0771}
S.~Samuroff,$^{1,3}$\orcidlink{0000-0001-7147-8843} and
R.~Mandelbaum$^{1,2}$\orcidlink{0000-0003-2271-1527}
\\
$^{1}$McWilliams Center for Cosmology, Department of Physics, Carnegie Mellon University, Pittsburgh, PA 15213, USA\\
$^{2}$NSF AI Planning Institute for Physics of the Future, Carnegie Mellon University, Pittsburgh, PA 15213, USA\\
$^{3}$Department of Physics, Northeastern University, Boston, MA, 02115, USA\\
}

\begin{document}

\maketitle
\begin{abstract}

In cosmology, we routinely choose between models to describe our data, and can incur biases due to insufficient models or lose constraining power with overly complex models.
In this paper we propose an empirical approach to model selection that explicitly balances parameter bias against model complexity.
Our method uses synthetic data to calibrate the relation between bias and the $\chi^2$ difference between models. This allows us to interpret $\chi^2$ values obtained from real data (even
if catalogues are blinded) and choose a model accordingly.
We apply our method to the problem of intrinsic alignments -- one of the most significant weak lensing systematics, and a major contributor to the error budget in modern lensing surveys.
Specifically, we consider the example of the Dark Energy Survey Year 3 (DES Y3), and compare the commonly used nonlinear alignment (NLA) and tidal alignment \& tidal torque (TATT) models. The models are calibrated against bias in the $\omegam - S_8$ plane.
Once noise is accounted for, we find that it is possible to set a threshold $\Delta \chi^2$ that guarantees an analysis using NLA is unbiased at some specified level $N\sigma$ and confidence level.
By contrast, we find that theoretically defined thresholds (based on, e.g., $p-$values for $\chi^2$) tend to be overly optimistic, and do not reliably rule out cosmological biases up to $\sim 1-2\sigma$.
Considering the real DES Y3 cosmic shear results, based on the reported difference in $\chi^2$ from NLA and TATT analyses, we find a roughly $30\%$ chance that were NLA to be the fiducial model, the results would be biased (in the $\omegam - S_8$ plane) by more than $0.3\sigma$. 
More broadly, the method we propose here is simple and general, and requires a relatively low level of resources. We foresee applications to future analyses as a model selection tool in many contexts.
\end{abstract}

\begin{keywords}
methods: statistical -- cosmology: observations -- cosmological parameters -- gravitational lensing: weak
\end{keywords}


\section{Introduction}\label{sec:intro}

Modern cosmology is an increasingly high-dimensional problem. Although the standard cosmological model itself is relatively simple, containing only five or six free parameters, it cannot, in general, be constrained in isolation. One must rely on measurements on real data, which can contain any number of additional features resulting from non-cosmological processes. It is necessary to include models for such systematics in any cosmological inference, and to marginalise over their parameters. Contemporary weak lensing analyses (see, e.g., \citealt{heymans13,sv-cosmicshear,DLS,hildebrandt17,y1-cosmicshear,hikage19, hamana20,asgari20,loureiro21,y3-cosmicshear1}; \citealt*{y3-cosmicshear2};  \citealt{y3-cosmicshearcls}) typically have around $15-30$ free parameters, the majority of which are related to measurement uncertainties.
This picture is unlikely to change in the coming years. Indeed, as we move into the era of Stage IV surveys (\citealt{lsst2019,roman,euclid}), the unprecedented statistical power of these new data sets carries an increasing sensitivity to systematics.

Some systematic uncertainties can be modelled pretty accurately given our prior knowledge of their nature; for instance PSF modelling error \citep{jarvis21} and shear measurement biases \citep{heymans06,bridle10,mandelbaum15}.
In most cases, however, there is a relative lack of prior knowledge about the magnitude and/or scale dependence of the effects being modelled. Some examples include the impact of baryonic feedback \citep{osato15,chen22,troster22}, nonlinear structure formation (and the impact of neutrinos on it; \citealt{saito08,bird12,mead21,knabenhans21}) and galaxy bias \citep{desjacques18,simon18,pandey20}. 
Here, there is clearly an argument for using the most sophisticated (physically motivated) model available. This is the safest way to avoid bias due to model insufficiency. That said, extra free parameters do potentially carry a cost in terms of constraining power. They can also worsen projection effects, which complicate the interpretation of projected parameter constraints (see \citealt{joachimi21, krause21}). 
The ideal approach, then, would be to select a model that balances the two: complex enough to avoid bias, but not more complex than is needed to describe the data. 

Model selection methods are widely used in cosmology, often seeking to answer the question of whether introducing new parameters to cosmological models is justified by the data. Some of the most common tools for this are $\chi^2$ tests, the Akaike Information Criterion (AIC), the Bayesian Information Criterion (BIC), the Deviance Information Criterion (DIC) and  Bayes ratios (see e.g. \citealt{liddle06,liddle07,trotta07,trotta08,Kerscher19}). A characteristic of how all these statistics have been used is that they are interpreted using threshold statistical values, derived in terms of the theoretical properties of the model, e.g. the Jeffreys scale for Bayes ratios. They have also been most commonly applied to compare how well cosmological models fit the data post-analysis, rather than actively being used to select elements of the analysis in the blinded stages.
By contrast, the process for choosing the fiducial model for an analysis typically does not make use of model comparison statistics at all. Rather, we tend to rely on generating and analysing simulations (either analytic or numerical) containing various forms of unmodelled systematics (e.g., \citealt{krause17,krause21,joachimi21}). This approach works, but does heavily depend on the ability to create realistic mocks. 
It is also important to notice that any model selection method will typically have a number of subjective choices built into them, e.g., whether to compare data vectors or perform likelihood inference, and what cutoff to use for decision-making. This is also true, to an extent, for the method we will present in this paper. That said, our method has the feature that the decision-making happens in well-defined places and has a well-defined interpretation connected to parameter biases (e.g. selecting a tolerable bias level and a confidence interval), as we will see in the following sections.

One of the most significant sources of systematic uncertainty in weak lensing is an effect known as intrinsic alignment (IA; \citealt{troxel15,joachimi15,kirk15,kiessling15}). 
IAs are coherent galaxy shape alignments that are not purely due to lensing, but rather to the interactions with the local and large-scale gravitational field.
Although in essence an astrophysical effect, IA correlations appear on much the same angular scales as cosmological ones, and it can be very difficult to disentangle the two. 
They are not universal, in the sense that they depend significantly on the particular galaxy sample (colour, luminosity, satellite fraction and redshift distribution; e.g. \citealt{johnston19}), and also the details of the shape measurement \citep{singh16}. To add to the problem, unlike, for example, photometric redshift error or shear bias, one cannot easily derive tight priors on IAs using simulations or external data.
Some physically-motivated IA models that have been developed in the last two decades include the linear alignment (LA) model \citep{catelan01,hirata04,hirata10} which, as the name suggests, assumes a linear relationship between galaxy shapes and the local tidal field; an empirical modification of this, known as the nonlinear alignment model (NLA; \citealt{hirata07,bridle07}), which is now one of the most common IA models in contemporary weak lensing; and in recent years, the tidal alignment and tidal torquing model (TATT; \citealt{blazek15,blazek17}), which has provided a slightly more complex alternative to NLA. Based on perturbation theory, TATT includes additional terms that are quadratic in the tidal field, intended to encapsulate the processes driving IAs in spiral galaxies, and also additional terms that are designed to enable better IA modelling on smaller (but still 2-halo) scales.

In this paper we propose a new model selection method, which uses the real data. The general idea is to run two competing models on the blinded data, and compare them using statistical metrics. Here we explore two convenient metrics: the difference in the best $\chi^2$ per degree of freedom, $\Delta \chi^2_{(\textrm{df})} = \Delta \chi^2/\Delta \mathrm{df}$, and the Bayes ratio $R$. We show that, for the method we are proposing, $\Delta \chi^2_{(\textrm{df})}$ is a very useful metric to perform model selection ($R$ is less so, for reasons discussed in Section \ref{sec:results:noisy} and Appendix \ref{sec:appendixBayes}). To allow us to interpret the $\Delta \chi^2_{(\textrm{df})}$, we use simulated data to calibrate its relation to biases in parameter space due to model insufficiency. It is this process, of running a set of simulations and measuring parameter bias as a function of observable metrics that we refer to as ``calibrating the bias-metric relation" in the following sections. The full details on how to perform this calibration are outlined in Section \ref{sec:method}. 
This approach can, in principle, be applied to any type of data and/or systematics. However, in what follows, we apply it the specific scenario of choosing an intrinsic alignment model for a cosmic shear analysis.

The paper is structured as follows: we describe the theoretical modelling of the cosmic shear two-point data vector in Section \ref{sec:theory}. In Section \ref{sec:data} we describe how the synthetic cosmic shear data is generated, including the choice of IA scenarios. 
The ingredients and steps for the model selection method are described in Section \ref{sec:method}. Our results when applying our method to the problem of IAs in the Dark Energy Survey Year 3 (DES Y3) are presented in Section \ref{sec:results}. Finally, in Section \ref{sec:conclusions} we summarise our findings and their significance in the context of the field.


\section{Theory \& Modelling}\label{sec:theory}
 
We carry out our analysis in the context of the flat $\Lambda$CDM cosmological model. The cosmological parameters are
$\left\{ \Omega_{\rm m}, \Omega_{\rm b}, h_0, A_{\rm s}, n_{\rm s}, \Omega_{\rm \nu}h^2 \right\}$,
where $\Omega_{\rm m}$ is the density parameter for matter, and $\Omega_{\rm b}$ the equivalent for baryons; $h_0$ is the dimensionless Hubble constant; $A_{\rm s}$ and $n_{\rm s}$ are the amplitude and slope of the primordial curvature power spectrum at a scale of $k=0.05$ Mpc$^{-1}$ respectively; and $\Omega_{\rm \nu}h^2$ is the neutrino mass density parameter. We assume three degenerate massive neutrino species, following \citet{krause21}. 
We discuss the nuisance parameters of our analysis in the following sections. Prior choices are further described in Appendix \ref{sec:appendixA}.

\subsection{Modelling Cosmic Shear}\label{sec:cosmicshear}

The impact of gravitational lensing along a particular line of sight is determined by two quantities, known as convergence and shear.
The \emph{convergence} $\kappa$ term of the weak lensing transformation describes how much a galaxy on a particular line of sight is distorted due to intervening large scale structure. 
It is defined as the weighted mass overdensity $\delta$, integrated along the line of sight to the distance of the source $\chi_\mathrm{s}$:
\begin{equation}
    \kappa\left(\boldsymbol{\theta}\right)= 
    \int_{0}^{\chi_{\mathrm{s}}}\mathrm{d}\chi\,W(\chi)\delta(\boldsymbol{\theta},\chi),\label{eq: convergence}
\end{equation}
where $\boldsymbol{\theta}$ is the angular position at which the source is observed. The kernel $W(\chi)$, defined in Eq.~(\ref{eq: lensing kernel W}), is sensitive to the relative distances of the source and the lens. It is this geometrical term that makes cosmic shear sensitive to the expansion history of the Universe.

The two-point cosmic shear correlations $\xi_\pm(\theta)$ are obtained by decomposing $\kappa$ into $E$- and $B$-mode components  \citep{CrittendenEB, Schneider_vanW_Mell_2002}. For two redshift bins $i$ and $j$, they can be written in terms of the convergence power spectrum $C_{\kappa}(\ell)$ at an angular wavenumber $\ell$ as
\begin{align} 
    \xi_{+}^{ij}(\theta) &=  \sum_{\ell}\frac{2\ell+1}{4\pi}G_{\ell}^{\pm}\left(\cos\theta\right)\left[C_{\kappa,EE}^{ij}(\ell)+C_{\kappa,BB}^{ij}(\ell)\right], \label{eq: xip Hankel transform}\\ 
    \xi_{-}^{ij}(\theta) &= \sum_{\ell}\frac{2\ell+1}{4\pi}G_{\ell}^{\pm}\left(\cos\theta\right)\left[C_{\kappa,EE}^{ij}(\ell)-C_{\kappa,BB}^{ij}(\ell)\right], \label{eq: xim Hankel transform}
\end{align}
\noindent
where the functions $G^\pm_\ell(x)$ are calculated from Legendre polynomials $P_\ell(x)$ and averaged over angular bins (see Eqs.~19 and 20 in \citealt{krause21}).

Assuming the Limber approximation \citep{Limber53, Limber_LoVerde2008}, the 2D convergence power spectrum $C^{ij}_\kappa(\ell)$ is related to the 3D matter power spectrum as: 
\begin{equation}\label{eq: Limber}
C_{\kappa}^{ij}(\ell)=\int_{0}^{\chi(z_\textrm{max})} \mathrm{d}\chi\frac{W^{i}(\chi)W^{j}(\chi)}{\chi^{2}}P_\delta \left(\frac{\ell+0.5}{\chi},z(\chi)\right),
\end{equation}
\noindent
where $P_\delta(k,z)$  is the nonlinear matter power spectrum and the lensing weight is:
\begin{equation}\label{eq: lensing kernel W}
W^{i}(\chi)=\frac{3H_{0}^{2}\Omega_{\mathrm{m}}}{2c^{2}}\frac{\chi}{a(\chi)}\int_{\chi}^{\chi_{\mathrm{H}}}\mathrm{d}\chi'\,n^{i}\left(z(\chi')\right)\frac{\mathrm{d}z}{\mathrm{d}\chi'}\frac{\chi'-\chi}{\chi'},
\end{equation}
\noindent
with the source galaxy redshift distribution $n^i(z)$ normalised to integrate to 1, and $\chi_\text{H}$ the horizon distance.
We follow \citet{krause21}, and model $P_{\delta}$ using a combination of \blockfont{CAMB} \citep{lewis00} for the linear part, and \blockfont{HaloFit} \citep{takahashi12} for nonlinear modifications.
In theory, the power spectra of convergence, $C_\kappa$, and cosmological shear, $C_\gamma$, are the same, and can be modelled fairly simply as described in Eq.~\eqref{eq: Limber}. In practice, however, $\xi_\pm$ measurements are sensitive not only to the pure cosmological shear, but also to additional correlations due to, e.g., intrinsic alignments. In the presence of IAs, the convergence spectra in Eqs.~\eqref{eq: xip Hankel transform} and~\eqref{eq: xim Hankel transform} are replaced by $C_\gamma$, the calculation of which we come to in Section \ref{sec:IAmodeling}.

\subsection{Modelling Intrinsic Alignments}\label{sec:IAmodeling}

In general terms, the impact of intrinsic alignments (IAs) can be thought of as adding a coherent additional component to each galaxy's shape. That is, the observed ellipticity is $\gamma^{\rm obs} = \gamma^{\rm G} + \gamma^{\rm I} + \epsilon_{SN}$, or the sum of a shear due to cosmological lensing, an IA-induced shear, and a random shape noise component. Although the latter is typically dominant for any single galaxy, it cancels when the ellipticity is averaged across a large population of galaxies. At the level of angular correlation functions, one has:

\begin{equation}
    C^{ij}_{\gamma}(\ell) = C^{ij}_{\kappa}(\ell) + C^{ij}_{\rm GI}(\ell) + C^{ji}_{\rm GI}(\ell) + C^{ij}_{\rm II}(\ell).
\end{equation}

\noindent
The first term, $C_\kappa$, is the auto-correlation of cosmological lensing, and is defined in Eq.~\eqref{eq: Limber}. The intrinsic-intrinsic contribution is referred to as the II term, and arises from galaxies that are spatially close to one another. The intrinsic-shear cross-correlation is known as the GI term, and emerges from the fact that galaxies at different distances along the same line of sight can either be lensed by, or experience gravitational tidal interaction with, the same large scale structure. 

Again assuming the Limber approximation, the angular power spectra can be written as

\begin{equation}\label{eq: CII}
    C^{ij}_{\rm II} (\ell) = \int_0^{\chi_{H}} \mathrm{d}\chi \frac{n^i(\chi) n^j(\chi)}{\chi^2}  P_{\rm II}\left ( k=\frac{\ell+0.5}{\chi}, \chi \right )
\end{equation}

\noindent
and
\begin{equation}\label{eq: CGI}
    C^{ij}_{\rm GI} (\ell) = \int_0^{\chi_{H}} \mathrm{d}\chi \frac{W^i(\chi) n^j(\chi)}{\chi^2}  P_{\rm GI}\left ( k=\frac{\ell+0.5}{\chi}, \chi \right ).
\end{equation}

\noindent
Given Eqs.~\eqref{eq: Limber}, \eqref{eq: CII} and \eqref{eq: CGI}, we have the ingredients to use Eqs.~\eqref{eq: xip Hankel transform} and \eqref{eq: xim Hankel transform} to predict the observable $\xi_\pm$.
Note that the GI and II power spectra are model dependent. Indeed, how one calculates them is a significant analysis choice in any cosmic shear analysis. In the sections below we describe the two model choices explored in this work.

\subsubsection{TATT Model}\label{sec:theory:tatt}

The tidal alignment and tidal torque model (TATT; \citealt{blazek17}) is based on nonlinear perturbation theory, which is used to expand the field of intrinsic galaxy shapes $\gamma^{\rm I}$ in terms of the tidal field $s$ and the matter overdensity $\delta$. 
Whereas $\delta$ is a scalar at all points in space, $\gamma^{\rm I}$ and $s$ are $3\times 3$ matrices, defining an  ellipsoid in 3D space.
Although in principle the expansion could be extended to any order, our implementation includes terms up to quadratic in the tidal field:
\begin{equation}
    \gamma^{\rm I}_{ij} = C_1 s_{ij} + C_2 \sum_k s_{ik}s_{kj} + b_{\rm TA} C_1 \delta s_{ij},
\end{equation}
where $C_1,C_2$ and $b_{\rm TA}$ are free parameters. This leads to the power spectra:

\begin{equation}\label{eq:tatt_gi}
    P_{\rm GI} = C_1 P_\delta + b_{\rm TA} C_1 P_{0|0E} + C_2 P_{0|E2},
\end{equation}

\begin{multline}\label{eq:tatt_ii}
    P^{\rm EE}_{\rm II} = C^2_1 P_\delta + 2 b_{\rm TA} C^2_1  P_{0|0E} + b^2_{\rm TA} C^2_1 P_{0E|0E} \\ + C^2_2 P_{E2|E2} + 
    2C_1C_2 P_{0|E2} + 2b_{\rm TA}C_1 C_2 P_{0E|E2},
\end{multline}

\begin{equation}\label{eq:tatt_ii_bb}
    P^{\rm BB}_{\rm II} = b^2_{\rm TA} C^2_1P_{0B|0B} + C^2_2P_{B2|B2} + 2b_{\rm TA} C_1 C_2 P_{0B|B2}.
\end{equation}

\noindent
The various subscripts to the power spectra indicate correlations between different order terms in the expansion of $\gamma^\mathrm{I}$. They can all be calculated to one-loop order as integrals of the linear matter power spectrum over $k$ (see \citealt{blazek17} for the full definitions).
As can be seen here, the TATT model predicts both $E$- and $B$-mode II contributions. These are propagated to separate $E$- and $B$-mode angular power spectra, which enter $\xi_\pm$ in Eqs.~\eqref{eq: xip Hankel transform} and \eqref{eq: xim Hankel transform}.  
The amplitudes are defined, by convention, as:

\begin{equation}\label{eq:tatt_a1}
    C_1(z) = -A_1  \frac{\bar{C}_1 \rho_{\rm c} \omegam}{D(z)} \left ( \frac{1+z}{1+z_0} \right )^{\eta_1},
\end{equation}

\begin{equation}\label{eq:tatt_a2}
    C_2(z) = 5 A_2  \frac{\bar{C}_1 \rho_{\rm c} \omegam}{D^2(z)} \left ( \frac{1+z}{1+z_0} \right )^{\eta_2}.
\end{equation}

\noindent
The pivot redshift $z_0$ and the constant $\bar{C}_1$ are fixed to values of $z_0=0.62$ and $\bar{C}_1 = 5\times10^{-14} M_\odot h^{-2} \mathrm{Mpc}^2$. Again, this is a convention, such that $C_1(z)$ and $C_2(z)$ are roughly of the order of $1$ for a typical population of source galaxies. 
The power law term in $C_1(z)$ and $C_2(z)$ adds some flexibility to capture possible redshift evolution beyond what is already encoded in the model.
Our implementation of the TATT model then has five free parameters: $A_1,A_2,\eta_1,\eta_2,b_{\rm TA}$, which we allow to vary with wide flat priors $A_1,A_2,\eta_1,\eta_2 \in [-5,5]$, $b_{\rm TA} \in [0,2]$. This choice of uniformative priors is motivated by the fact that IAs are very sensitive to the properties of the galaxy population, making it very difficult to derive informative priors, and resulting in a lack of directly transferable constraints in the literature for the TATT model parameters.
Although intended to match up with different alignment mechanisms, in practice $A_1$ and $A_2$ capture any correlations that scale linearly and quadratically with the tidal field. The third amplitude $b_{\rm TA}$ is designed to capture the fact that galaxies over-sample densely populated regions (i.e., one cannot sample the $\gamma^{\rm I}$ field uniformly throughout the Universe).

For this work we use the DES Y3 implementation of TATT, within \blockfont{CosmoSIS} v1.6\footnote{\url{https://bitbucket.org/joezuntz/cosmosis/wiki/Home}; des-y3 branch of cosmosis-standard-library}  \citep{zuntz15}. 
The power spectra in Equations~\eqref{eq:tatt_gi}-\eqref{eq:tatt_ii_bb} (with the exception of the nonlinear matter power spectrum $P_\delta$) are generated using \blockfont{FAST-PT} v2.1\footnote{\url{https://github.com/JoeMcEwen/FAST-PT}} \citep{mcewen16,fang17}.

\subsubsection{NLA Model}\label{sec:theory:nla}

Although chronologically older and more commonly used, the nonlinear alignment model  (NLA; \citealt{bridle07}) is a subspace of TATT. Built on the assumption that galaxy shapes align linearly with the background tidal field, it predicts:

\begin{equation}
    P_{\rm GI} = C_1(z) P_\delta, \;\;\; P_{\rm II} = C^2_1(z) P_\delta,
\end{equation}

\noindent
with the amplitude $C_1(z)$  as defined in Eq.~\eqref{eq:tatt_a1} in our implementation. 
The NLA model as implemented here differs from its predecessor, the linear alignment model \citep{catelan01,hirata04,hirata10}, by the fact that $P_\delta$ in the above equations is the full nonlinear matter power spectrum (in our case generated using \blockfont{HaloFit}), not the linear version. Unlike the original formulation, our implementation of NLA also includes a power law redshift dependence in $C_1(z)$ to capture any additional evolution beyond the basic model (as in Eq.~\eqref{eq:tatt_a1} above). 
In total, our implementation of the NLA model has two free parameters, $A_1$ and $\eta_1$, which we vary with the priors given in Section~\ref{sec:theory:tatt}.  

\subsection{Other Nuisance Parameters \& Scale Cuts}

Both the TATT and NLA pipelines include free parameters for redshift error and residual shear bias. We adopt the same modelling as \citet{krause21}, giving us one $\Delta z$ and one $m$ parameter per redshift bin, or a total of eight nuisance parameters. Note however, that these parameters are prior dominated for Y3 cosmic shear-only chains, and so add relatively little to the effective dimensionality. For details about the priors see Appendix~\ref{sec:appendixA} and Table~\ref{table: priors}.
We also adopt the fiducial DES Y3 cosmic shear scale cuts (see \citealt{krause21} for an explanation of how these were derived).


\section{Creating and analysing the cosmic shear data vector}\label{sec:data}

In this section we describe how we generate mock data. This is necessary to calibrate the relation between bias in cosmological parameters and statistical metrics used for model comparison,
which is central to our method for model selection. Essentially we wish to create an ensemble of data vectors that span a useful range of bias in cosmological parameters and $\Delta \chi^2_{(\textrm{df})}$ (or $R$), allowing us to map out the relation between the two.
In Section~\ref{sec:data:gen_data} we focus on how to simulate the cosmological lensing terms (which depend on cosmology, not the IA model). Then in Section~\ref{sec:data:ia_sampling} we describe the IA terms (IA model-dependent). We explain how we incorporate noise into our analyses, and why it is necessary, in Section~\ref{sec:data:noise}. Finally, we describe our sampler choices in Section~\ref{sec:method:sampler}, and in particular an approximation using importance sampling that we use to accelerate the analysis of the noisy data vectors.

\subsection{Generating Mock Data}\label{sec:data:gen_data}

For a given set of input parameters, we generate a noiseless DES Y3-like cosmic shear data vector, $\boldsymbol{D}$, using the theory pipeline described in Section~\ref{sec:theory}. We assume the fiducial Y3 redshift distributions, as presented in \citet*{y3-sompz}. In all data vectors, the same input flat \lcdm~cosmology is used $(\omegam=0.29, \; \as = 2.38\times10^{-9}, \; \omegab = 0.052, \; h = 0.75, \; \ns = 0.99, \; \Omega_{\nu}h^{2} = 0.00053 )$. This corresponds to $\sigma_8=0.79$, $S_8=0.77$, where $S_8 \equiv \sigma_8\sqrt{\omegam/0.3}$.
We choose these to match the marginalised mean values from the DES Y1 $3\times2$pt chain used to generate IA samples (see Section~\ref{sec:data:ia_sampling} below). Note, however, that the exact values are not expected to affect our results. We also fix all the redshift and shear calibration nuisance parameters to zero when generating data vectors.

\begin{figure}
\includegraphics[width=\columnwidth]{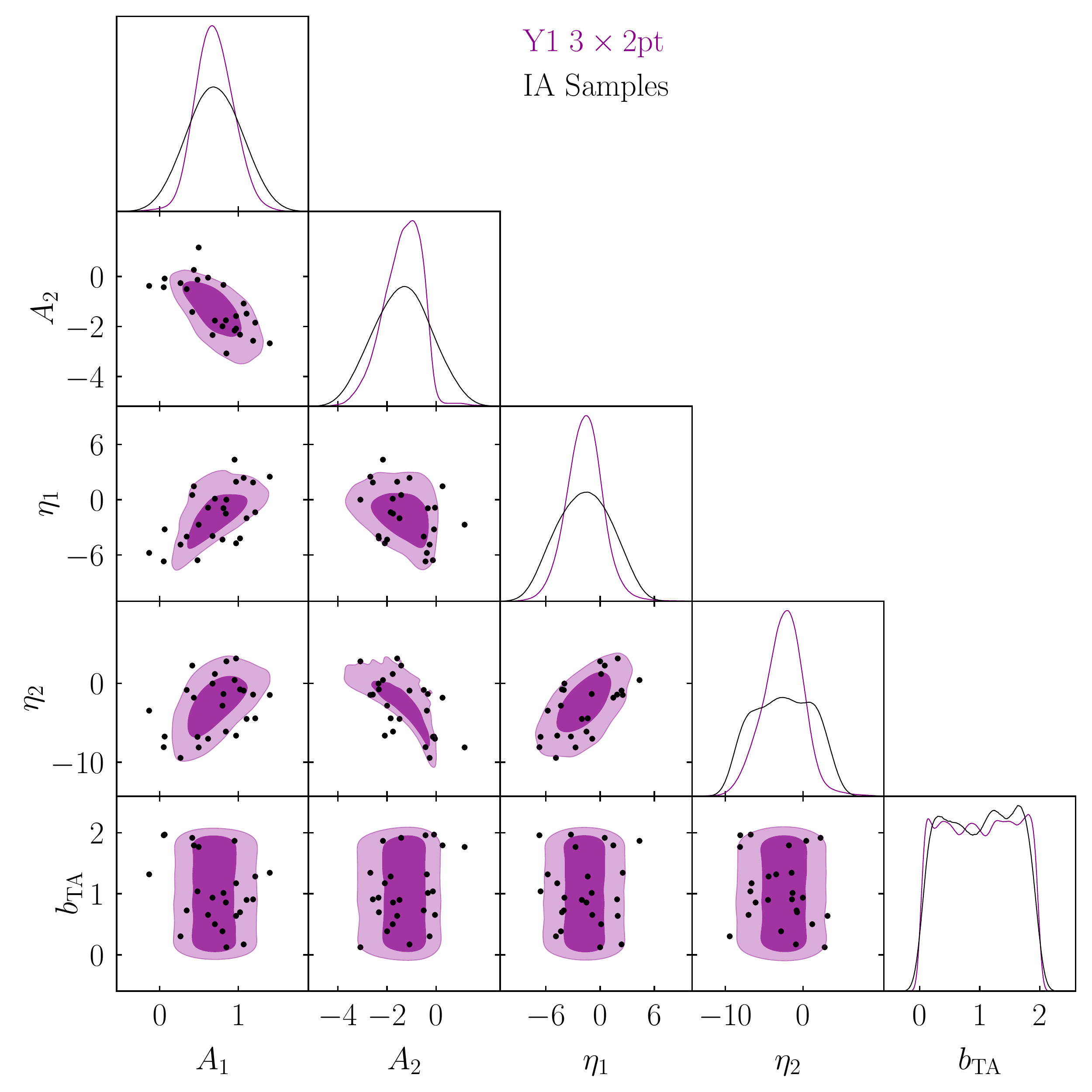}
\caption{An illustration of how we generate samples in IA model parameter space for this work. The purple contours show the $68\%$ and $95\%$ confidence levels from the TATT model analysis of the DES Y1 $3\times2$pt data \citep{y1-tatt}. Overlain (black points) are the IA samples we derive from this posterior probability distribution after marginalizing over all other parameters. On the diagonal, we show the Y1 marginal posterior (purple), and also the distribution of IA samples (black), both normalised to integrate to 1 over the prior range. As shown, the latter is slightly broader than we would obtain by drawing from the DES posterior distribution.
}
\label{fig:y1_contours}
\end{figure}

\subsection{Choosing IA Scenarios}\label{sec:data:ia_sampling}

When constructing simulated data vectors, it is important to remember that IA model parameters are not independent. The total GI+II intrinsic alignment component in a scenario with, e.g., $A_1=A_2=1$ is very different from one with $A_1=0.1,A_2=1$. As a consequence, it is possible for two sets of input IA parameters to give similar cosmological bias (when analysed with NLA), but quite different $\chi^2$ values. Specific combinations of TATT parameter values may  enhance or cancel out cosmological parameter bias, and so it is useful to sample the 5D TATT parameter space rather than scaling up individual parameters to explore the potential for cosmological parameter bias due to model insufficiency.  
Therefore, instead of a single mock data vector, we generate a set of 21 data vectors, all with the same cosmology, but with different possible IA scenarios.
The number of mock data vectors is an analysis choice. The more we generate, the better we cover the IA parameter space, but it also increases computational costs. We verified that 21 was a sufficient number of scenarios to have samples presenting low, medium and high bias in cosmological parameters, while still being reasonable in terms of computational cost (i.e. the chains to run). 

To do this, we follow the recipe set out in Section~2 of \citet{derose19}. Starting with the posterior from the DES Y1 $3\times2$pt TATT analysis (the purple contours in Figure \ref{fig:y1_contours}; \citealt{y1-tatt}), we evaluate the covariance matrix of the $N_p=5$ TATT parameters, and perform an eigenvalue decomposition. We then use Latin Hypercube sampling to generate $N_{\rm samp}$ samples, which are roughly evenly distributed in $N_p$ dimensional space. Finally, we use the eigenvalues/vectors to rotate and normalise those samples into the parameter space. The results are shown in Figure~\ref{fig:y1_contours}. The idea is that this provides a slightly broader coverage than could be obtained simply by drawing points from the joint posterior, while maintaining the correlations between parameters. In this way, we can cover a range of marginal cases, which are pessimistic, but still consistent with the data; this is useful, since for our purposes it is more important to span the range of plausible TATT model parameters than to preserve the statistics of the Y1 posterior exactly.

\subsection{Adding Noise}\label{sec:data:noise}

Since real measurements unavoidably include an (unknown) noise realisation, the calibration of the bias-metric relation is inherently a probabilistic problem (we will return to this point in Section~\ref{sec:results}; see the discussion there for details). For this reason, it is important that our simulations capture all sources of scatter in the data.

For each of our 21 IA scenarios, defined by a set of input TATT parameter values $\boldsymbol{\theta}_{\mathrm{IA},i}$, we have a set of noisy data vectors $\tilde{\boldsymbol{D}}_{i,j} = \boldsymbol{D}(\boldsymbol{\theta}_{\mathrm{IA},i}) + \boldsymbol{N}_j$, where noise realisation $\boldsymbol{N}_j$ is drawn from the covariance matrix, and is assumed to be independent of $\boldsymbol{\theta}_{\mathrm{IA},i}$. We use the final DES Y3 covariance matrix, which is analytic and includes a Gaussian shape noise and cosmic variance contribution, as well as higher order non-Gaussian and super-sample terms \citep{y3-covariance}. 
In total we generate 50 noise realisations, which we apply to each data vector.
This gives us a collection of 21 noiseless data vectors, and $21\times50=1050$ noisy ones.

\begin{figure*}
\includegraphics[width=2.\columnwidth]{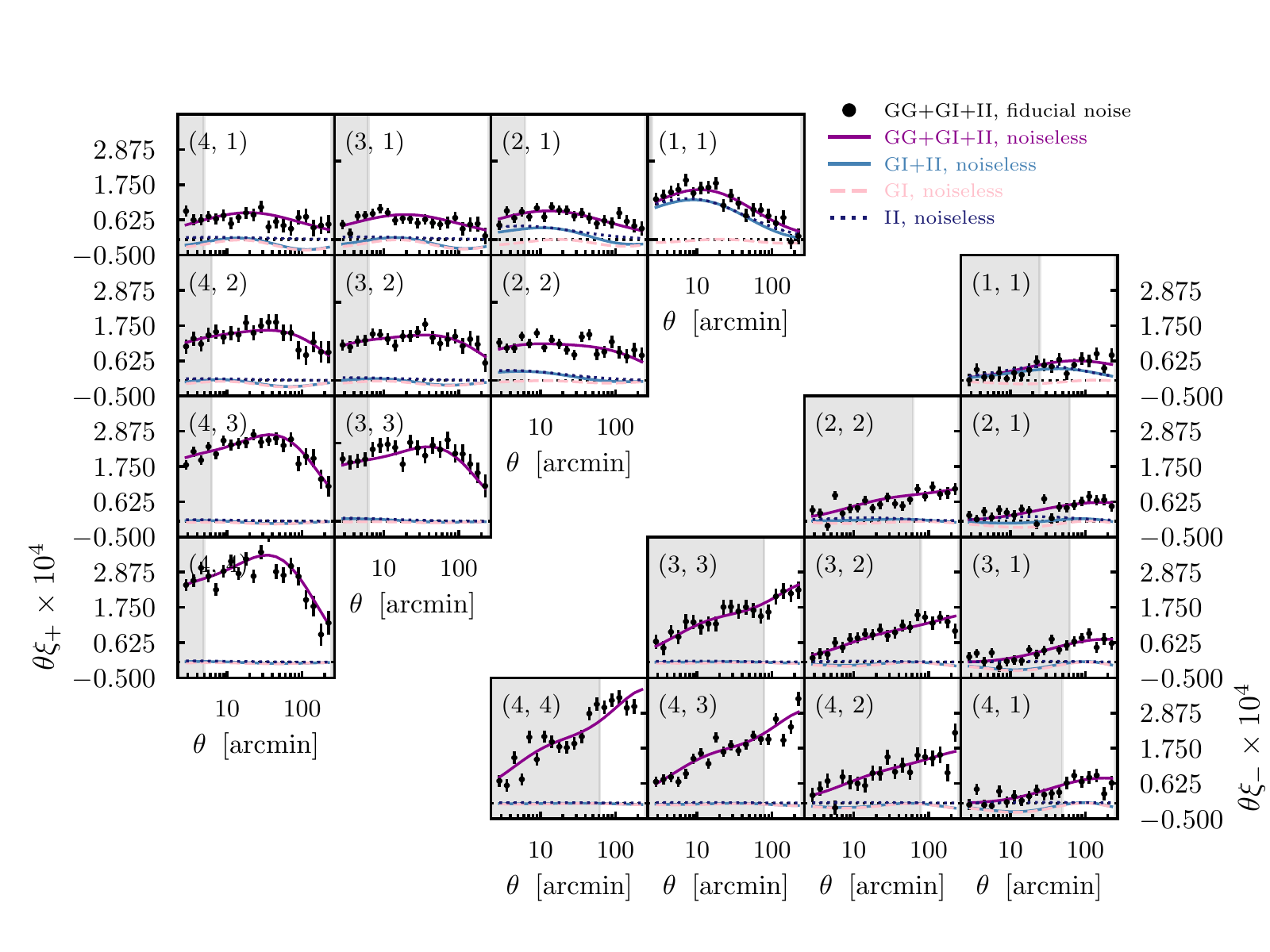}
\caption{An example of a noisy simulated data vector of the type used in this paper. Each panel shows a redshift bin combination (as labelled), and the upper and lower triangles show $\xi_+$ and $\xi_-$ respectively. In each panel we show the simulated cosmic shear data vector with fiducial noise (black points with error bars), as well as the noiseless version (smooth purple). 
We also show the GI and II intrinsic alignment components separately. For reference, the input IA parameters here correspond to the mean of the DES Y1 posterior discussed in Section~\ref{sec:data:ia_sampling} $(A_1=0.7,A_2=-1.36,\eta_1=-1.7,\eta_2=-2.5,b_{\rm TA}=1)$.
The grey bands represent the fiducial DES Y3 cosmic shear scale cuts, i.e., the scales removed from our analysis.
}
\label{fig:noisy_data_vec}
\end{figure*}

For testing, however, it is convenient to arbitrarily choose a single fixed noise realisation, which we refer to as \textit{fiducial noise}. 
Figure~\ref{fig:noisy_data_vec} shows an example DES Y3-like data vector, generated using the setup described above, with the fiducial noise realisation added. For this particular example, the input TATT parameters are the mean values from the Y1 posteriors in Figure~\ref{fig:y1_contours}. For reference, we also show the noiseless version (purple solid), as well as the separate (again, noiseless) IA contributions. 
Since the IA signal in the lowest bin ($1,1$) seems to dominate, one could reasonably ask whether we could simply focus on this part of the data vector for model selection. We ultimately decide against this for a few reasons. First, although the IA signal is strongest in bin $1,1$, we can see there is also non-negligible signal in the surrounding bins (e.g. $2,1$ and $3,1$). Indeed, during the DES Y3 analysis, a range of IA mitigation techniques were explored, including dropping the lowest auto bin correlations. For moderate TATT scenarios, it was found that this could not reliably eliminate cosmology biases, suggesting the IA contamination is not confined to the $1,1$ data vector. Additionally, the degree to which the low redshift II contribution dominates the IA signal is somewhat dependent on the input TATT parameters, and so the strength of this assumption varies in parameter space.

\subsection{Choice of Sampler}\label{sec:method:sampler}

\subsubsection{Nested Sampling}\label{sec:method:sampler:nested}

To sample the cosmic shear likelihood, we use the \blockfont{PolyChord} nested sampling algorithm \citep{handley15}, which generates estimates for the multidimensional posterior $\mathcal{P}(\boldsymbol{\theta}|\boldsymbol{D},M)$ and the evidence for a given model $\mathcal{Z}(\boldsymbol{D}|M)$ simultaneously.
This matches the DES Y3 choice, and has been validated in terms of both evidence and the contour size compared with a long monte carlo chain \citep*{y3-samplers}. We briefly explored the possibility of using \blockfont{MultiNest} \citep{feroz13}, which is conceptually similar, but significantly faster. Ultimately, however, we found that \blockfont{MultiNest} underestimates the width of the posteriors in all cases we tested (both NLA and TATT; see Appendix \ref{app:multinest_polychord}). It also gives inaccurate evidence values \citep*{y3-samplers}, which tend to skew towards preferring NLA. For these reasons, we did not pursue this. 

To obtain estimates for the best $\chi^2$, we use oversampled chains (i.e., output with $10\times$ the number of points as saved in the standard chains). This approach has been tested in the Y3 cosmic shear setup, and shown to give comparable results to running a likelihood maximiser \citep*{y3-cosmicshear2}. All sampling, as well as the modelling steps described in Section \ref{sec:theory}, are carried out using \blockfont{CosmoSIS}. 

\subsubsection{Importance Sampling}\label{sec:modelling:importance_sampling}

To assess the impact of data vector noise, in addition to nested sampling we also make use of importance sampling (IS; \citealt{neal98,todkar10} see also \citealt{lewis02,padilla19} for cosmology-specific applications). For each IA scenario $\boldsymbol{\theta}_{\mathrm{IA},i}$, we wish to estimate the shape and position of the $S_8-\omegam$ posterior, as well as the best fit and evidence. Running full chains for every combination of noise and IA scenario would be expensive, and IS provides a fast approximation. 

Say one wants to estimate the characteristics of a distribution $P$, over parameter $\theta$. 
One can estimate the mean of the function $f(\theta)$ under $P$ as:

\begin{equation}
\hat{f} = \int f(\theta) P(\theta) \mathrm{d}\theta.
\end{equation}

\noindent
This can be rewritten in terms of a second distribution $P_0$:

\begin{align}
\hat{f} = \int f(\theta) \frac{P(\theta)}{P_0(\theta)} P_0(\theta) \mathrm{d}\theta\\
\approx \sum_{\theta_i \sim P_0} f(\theta_i) w(\theta_i),
\end{align}

\noindent
where we have redefined the ratio of distributions as a weight $w=P/P_0$. The second line follows as a Monte Carlo estimate for the first, and the sum runs over values of $\theta$ drawn from $P_0(\theta)$.  
The equations above make no assumptions about Gaussianity, or about the nature of the distributions. To work well, however, it does require $P_0$ to be non-zero over the range of $\theta$ for which we wish to estimate $P$, and it works better in cases where the number of samples is large. Functionally, it also requires (a) that one has, or can generate, samples from $P_0$, and (b) that for any given $\theta$, one can evaluate both $P$ and $P_0$.  

For our application, $P_0=\mathcal{P}(\boldsymbol{\theta} | \boldsymbol{D}, M)$ is a reference posterior obtained from running a chain on the noiseless data vector $\boldsymbol{D}(\boldsymbol{\theta}_{\rm IA})$. As before, we use the $\times10$ oversampled \blockfont{PolyChord} output for this. The target distribution $P=\mathcal{P}(\boldsymbol{\theta} | \tilde{\boldsymbol{D}}, M)$ is the posterior we are trying to estimate, conditioned on a noisy data vector $\tilde{\boldsymbol{D}}$. With this setup, we can estimate $P$ for each noise realisation by simply iterating through the samples from $P_0$ and assigning each a weight equal to the ratio of the two posteriors.

In addition to the target posterior for each model, we also estimate the best $\chi^2$. For this, we create a high density pool of samples by merging all of our oversampled \blockfont{PolyChord}  chains (21 IA samples), in addition to a small number of additional chains run with a Y1 like covariance matrix. This gives us over a million points in parameter space per model. For each noisy data vector, we re-evaluate the likelihood at each point, and select the maximum. Given an estimate for the best $\chi^2$ from IS, and assuming a Gaussian likelihood, the Bayesian evidence can then be estimated as (see Section 3 \citealt{joachimi21a}):
\begin{equation}\label{eq:approx_evidence}
    \mathrm{ln}\mathcal{Z}(\tilde{\boldsymbol{D}_i}|M) = \mathrm{ln}\mathcal{Z}_0(\boldsymbol{D}|M) - 0.5 \left (\chi^2_{i} - \chi^2_0 \right ),
\end{equation}
\noindent
where index $i$ indicates a noise realisation, and $\mathcal{Z}_0$ and $\chi^2_0$ are the evidence and best $\chi^2$ obtained from a fiducial reference chain, which in our case are our noiseless chains.

To test the performance of our IS setup, we ran five additional \blockfont{PolyChord} chains at different noise realisations, once in the low bias regime and again in the intermediate ($\sim 1 \sigma$) bias regime. We verified that in all cases, our IS setup recovered the best $\chi^2$ as well as the shape and mean of the posteriors with a comparable level precision to a full chain. Our implementation is a slight modification of the code discussed in \citet*{weaverdyck22}, and will be available on release of that paper.


\section{Model Selection}\label{sec:method}

In this section, we define the components of our model selection method. In essence, we are proposing to calibrate the observed value of model comparison statistics against the probability of cosmological parameter bias. These quantities can be computed using noisy simulated data, but first they must be properly defined. To this end, we define how we quantify cosmological parameter bias in Section~\ref{sec:method:bias}. The metrics that we tested in search of a useful bias--metric relation are discussed in Section~\ref{sec:method:stats}.
In Section~\ref{sec:method:unphysical} we make considerations regarding unconverged samples.
We summarize our model selection method in Section~\ref{sec:method:recomendation}.

\subsection{Significance Level of Cosmological Parameter Biases}\label{sec:method:bias}

\begin{figure}
\includegraphics[width=\columnwidth]{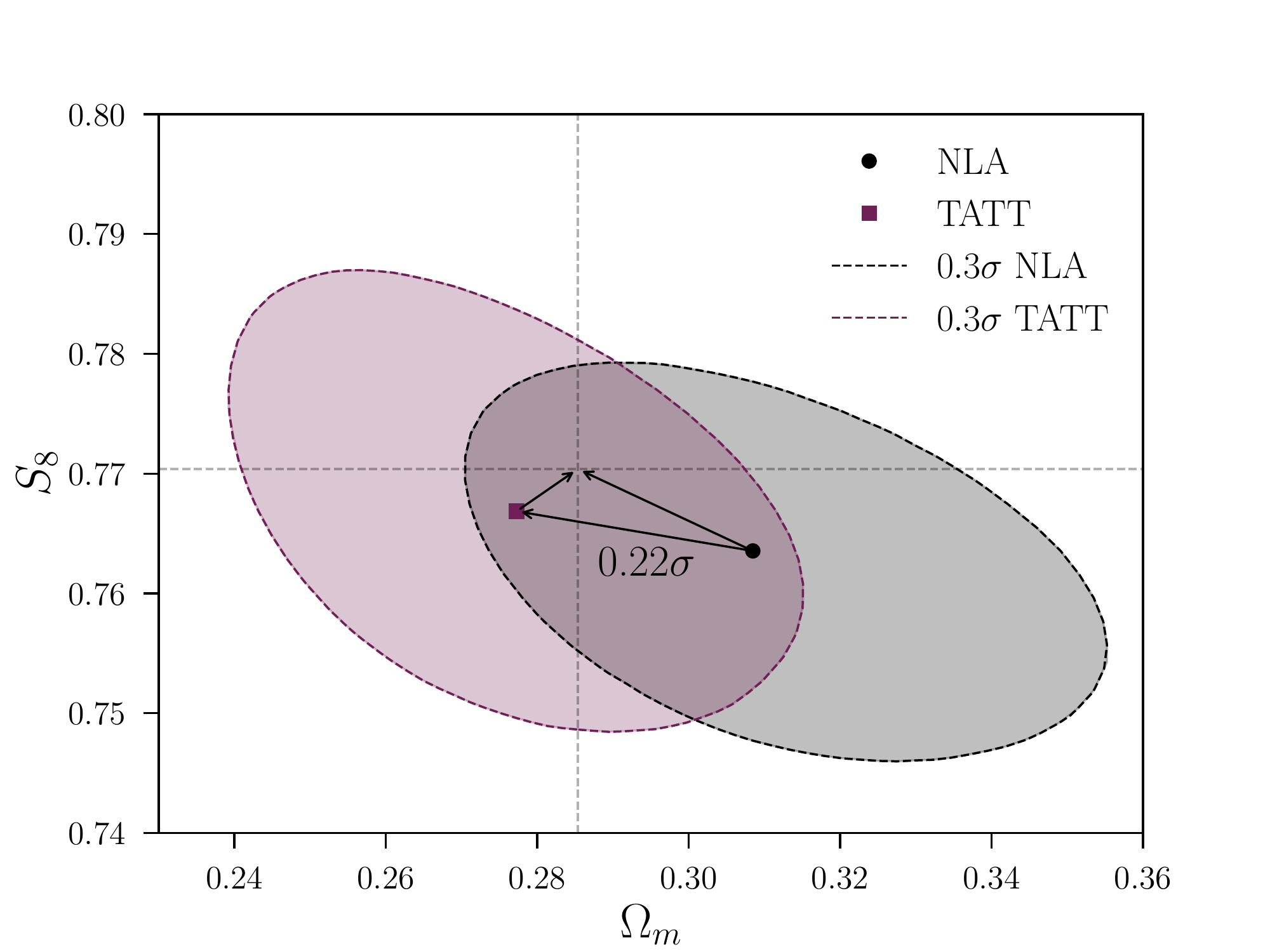}
\caption{An example of how cosmological parameter bias is defined for a given IA scenario and noisy data vector. The purple point and the dotted ellipsoid show the maximum likelihood and $0.3\sigma$ contour, obtained from the analysis of a noisy simulated data vector with the TATT model. The black is the same, but with the NLA model. The vector connecting the two peaks in the $\omegam-S_8$ plane defines our bias metric. Note that the TATT contour is slightly offset from the input parameter values (the dashed lines) due to noise and projection effects. It is for this reason that the relative separation, rather than the distance from the input, is the most appropriate bias definition.
}
\label{fig:ex_bias}
\end{figure}

Now that we have a set of noisy data vectors,
the next step is to fit them with all parameters free. We analyse each data vector twice, fitting to our full set of cosmological and nuisance parameters, but in one case using TATT, and in the other NLA. 
We then define bias as the distance between the peak of the marginalised posteriors in the $S_8 - \omegam $ plane. Figure~\ref{fig:ex_bias} illustrates this for a particular simulated data vector. 
In brief, the algorithm works by evaluating the Euclidean distance between the peaks of the two posteriors, $\mathcal{P}_1$ and $\mathcal{P}_2$ in $S_8-\omegam$ space. 
It then sequentially computes the confidence ellipses of $\mathcal{P}_1$ at different $\sigma$ levels, and finds a value of $N_\sigma$ that minimises the distance between the ellipse and the peak of $\mathcal{P}_2$.
Note that this is the same recipe used in \citet{krause21}.
The choice of $S_8$ and \omegam~as the parameters of interest comes simply from the fact that these are the cosmological parameters best constrained by DES. One could conceivably use a more complicated separation metric that is sensitive to the full parameter space, along the lines of those used for assessing tensions between data sets (e.g. \citealt*{y3-tensions}). For our purposes, however, we follow \citet{krause21} and consider the simpler 2D metric to be sufficient.

We use noisy simulated data, as described in Section~\ref{sec:data:noise} -- this is an important feature of our analysis, and it is necessary to allow us to meaningfully interpret our statistical metrics.
Therefore, the relative separation of the two posteriors is a more useful quantity than the distance from the input values of $S_8$ and \omegam. 
The $0.22\sigma$ value shown in Figure~\ref{fig:ex_bias} is assessed relative to the TATT posterior. This found to be more stable than assessing it relative to the NLA posterior, particularly in relatively extreme IA scenarios where the NLA posteriors are significantly shifted and can be distorted by prior edge effects. 

Finally, it is implicit in the above that marginalised TATT constraints represent correct results by which to measure bias. That is, when we refer to \emph{bias}, we are in fact talking about \emph{bias in the cosmological model when assuming the NLA model, with respect to what we find when assuming the TATT model}. Although this is clearly reasonable (since our data were created using TATT), marginal contours can be subject to projection effects. Indeed, since some of the TATT parameters are relatively poorly constrained by shear-shear analysis alone, the two IA models cannot be assumed to experience projection effects to the same degree. We test this in Appendix~\ref{sec:appendixA}, and find projection offsets between TATT and NLA at the level of $0.1\sigma$. This is well below the threshold of $0.3\sigma$ used for this work, and is thus unlikely to significantly affect our results.

\subsection{Model Comparison Statistics}\label{sec:method:stats}

We investigate two commonly used test statistics, the $\chi^2$ difference and the Bayes ratio. We show in Section \ref{sec:results} that the former is more robust against noise than the latter, and is therefore a more useful metric for the method we are proposing.

\subsubsection{$\chi^2$ Difference Tests}\label{sec:method:stats:chi2} 

When dealing with nested models (i.e., where one model is a subspace of the other, as in the case with NLA and TATT), the difference in the best $\chi^2$ that can be achieved by each model on the same data is a convenient statistic for model selection \citep{steiger85,rigdon99,engel03,andrae10}.

The $\chi^2$ difference metric is defined as the difference in the best $\chi^2$ values of the parameter fits when assuming the two IA models, divided by the difference in their numbers of degrees of freedom ($df$):

\begin{equation}\label{eq:deltachi2}
    \Delta\chi^2_{(\textrm{df})} = \frac{\chi^2_s - \chi^2_l}{d f_s - d f_l},
\end{equation}

\noindent
where \textit{s} indicates the \textit{smaller} model (the one with fewer free parameters and therefore more degrees of freedom; NLA in our example), and \textit{l} denotes the model with more parameters, \textit{larger}, and so fewer degrees of freedom (TATT in our case)\footnote{Note that it is this quantity, weighted by the difference in $df$, that we refer to as $\Delta \chi^2_{(\textrm{df})}$ throughout the paper, and not the simple $\chi^2_s - \chi^2_l$ difference. This allows us to briefly compare with theoretical cut-offs in Section \ref{sec:results:noiseless}. For practical purposes, however, this is not strictly necessary -- one could just as easily calibrate the raw difference.}. Note that the point estimate to use for the ``best $\chi^2$" here is somewhat subjective. For a given chain, we use the value closest to the peak of the multidimensional posterior. One could also use the maximum likelihood, although in practice this tends to be a slightly noisier quantity.

In the limit that the additional parameters in the larger model have no impact on the quality of the fit, the metric is exactly zero: $\Delta \chi^2 = \chi^2_s - \chi^2_l = 0$. Very small $\Delta \chi^2_{(\textrm{df})}$ values can therefore be taken as evidence that the smaller model is sufficient, given the data. In practice, however, this is an unlikely outcome, as extra parameters will typically allow the model more flexibility. Under the null hypothesis that the two models $s$ and $l$ both adequately fit the data, the value of the numerator $\chi^2_s - \chi^2_l$ is $\chi^2$-distributed with $df_\text{diff} = d f_s - d f_l$ degrees of freedom, and the expectation value is $\langle \Delta \chi^2 \rangle \sim 1$ \citep{wilks38}. One can interpret larger $\Delta \chi^2$ using the corresponding \textit{p-values} to quantify the degree to which the data appear to favour the larger model. 
As we will discuss in Section~\ref{sec:results}, 
however, for our purposes it is more useful to focus on the observed relation between $\Delta \chi^2_{(\textrm{df})}$ and parameter bias than on formal statistical thresholds. That is, we propose to use $\Delta \chi^2_{(\textrm{df})}$ as an empirical tool, which requires calibration using simulated likelihood analyses for any given problem. This way, we are also free from other  assumptions behind the standard use of the $\Delta\chi^2_{(\textrm{df})}$ metric -- for example, it formally requires nested models whereas an empirical calibration would not. 
Note that our approach here is functionally similar, but motivated slightly differently, to the calibration of Posterior Predictive Distribution (PPD) $p-$values for internal consistency testing, as implemented in \citet*{y3-ppd}.

In principle, $\Delta \chi^2_{(\textrm{df})}$ is prior-independent. In Bayesian inference, however, the prior typically controls the regions of parameter space that can be explored, and so restrict the values of $\chi^2$ that can be attained. In practice, this is only an issue if the likelihood peaks outside the prior bounds (which is, in any case, usually a red flag).

One other point to remember is that, although in an ideal case with well-constrained parameters one extra parameter constitutes one fewer degree of freedom, in practice this is often not true. In such cases, one can calculate the \emph{effective} degrees of freedom (see~\citealt{raveri19}). With the fiducial DES Y3 cosmic shear setup (minus the shear ratios), the effective degrees of freedom for TATT and NLA are 222 and 224 respectively, giving $df_\text{diff}=2$ (compared with $df_\text{diff}=3$ from simple parameter counting; see \citealt*{y3-cosmicshear2} Table~III).

In Section~\ref{sec:results}, we also briefly consider two other likelihood-based metrics: the Akaike Information Criterion (AIC; \citealt{akaike73}) and the Bayesian Information Criterion (BIC; \citealt{schwarz78}). Although these statistics have very different theoretical underpinnings (see \citealt{liddle07}),  they are similar in form, and can be conveniently reformulated as thresholds in $\Delta \chi^2_{(\textrm{df})}$. As with $p-$value cut-offs, however, they are seen to be relatively under-cautious in separating high- and low-bias scenarios (see Section~\ref{sec:results} and Figure~\ref{fig:bias_results}).  

\subsubsection{Bayes Ratio}\label{sec:method:stats:bayes}

The Bayesian evidence ratio, or Bayes ratio $R$ \citep{jeffreys68,kass95}, is a slightly more complicated alternative to $\Delta \chi^2_{(\textrm{df})}$. It is defined as the probability of measuring a data vector $\boldsymbol{D}$ assuming a given model $M_1$, divided by the probability of measuring the same data $\boldsymbol{D}$ for a second model $M_2$: 
\begin{equation}
\label{eq:bayesr}
    R \equiv \frac{\mathcal{Z}(\boldsymbol{D}|M_1)}{\mathcal{Z}(\boldsymbol{D}|M_2)}.
\end{equation}
Here, $\mathcal{Z}(\boldsymbol{D}|M)$ is the Bayesian evidence, which can be obtained marginalising over all the model parameters $\boldsymbol{\theta}$:
\begin{equation}
\label{eq:evidence}
    \mathcal{Z}(\boldsymbol{D}|M) = \int \rmd \boldsymbol{\theta} \ \mathcal{L}(\boldsymbol{D} |\boldsymbol{\theta}, M) P(\boldsymbol{\theta} | M),
\end{equation}
where $\mathcal{L}(\boldsymbol{D} | \boldsymbol{\theta},  M)$ is the likelihood, and $P(\boldsymbol{\theta} | M)$ is the prior, both assuming a particular model. The Bayes ratio is typically interpreted using the Jeffreys scale \citep{jeffreys68}, which defines ranges of values that match up to labels (e.g., ``strong evidence", ``substantial evidence", etc.).

Note that $R$ and $\Delta \chi^2_{(\textrm{df})}$ are not independent from one another (indeed the latter approximates the former under certain assumptions; see \citealt{bishop06,marshall06}). It is important, then, to be careful when seeking to combine information from the two.

Evidence ratios have been widely used in cosmology, both for comparing different data sets under the same model (i.e., as a tension metric; \citealt{marshall06}; \citealt*{y3-tensions}), and for model comparison on the same data \citep*{liddle06,kilbinger10,y3-cosmicshear2}.
It is worth bearing in mind that the formulation in the two contexts is slightly different. In the former case there is explicit prior dependence, which motivates the use of statistics such as Suspiciousness (see e.g.~\citealt*{y3-tensions} Section~4.2). The version commonly used for model selection, on the other hand, should be independent of the choice of priors, at least in the limit that (a) the models are nested and (b) the priors on the extra parameters are wide compared with the likelihood.

Since cosmological analyses involve a large number of free parameters, computing the Bayesian evidence requires integrating a probability distribution over a high number of dimensions. A common way to calculate it is while producing the posterior distributions, using nested sampling \citep{Skilling:2006}. The precision required from the sampler in order to compute reliable Bayesian evidences, however, often makes the sampling time very long. 
We choose to use the \blockfont{PolyChord} nested sampling algorithm in this work \citep{handley15} -- although see Appendix \ref{app:multinest_polychord}, where we consider the feasibility of using \blockfont{MultiNest} \citep{feroz13} as a slightly faster alternative.

\subsubsection{Bias Probability}\label{sec:method:bias_probability}

The above quantities give us the basic tools for our model comparison. There is, however, a piece missing. As we mention in Section~\ref{sec:data:noise}, 
the calibration is inherently probabilistic. The model comparison metrics (both $\Delta\chi^2_{\rm (df)}$ and $R$),
as well as the offset between the NLA and TATT best fits, are somewhat sensitive to noise, and we do not know the true noise realisation in the data. We thus define a \emph{bias probability} $P$ for a particular bias tolerance $X$:
\begin{equation}
P(b>X\sigma | \Delta \chi^2_{\rm (df), obs} < \Delta \chi^2_{\rm (df), thr}) = \frac{N^{b>X}_{\rm samp}}{N^{b>X}_{\rm samp} + N^{b<X}_{\rm samp}}.
\label{eq:prob_threshold}
\end{equation}
\noindent
In words, $P$ is the probability of the bias in $S_8$--\omegam~being greater than $X\sigma$, \emph{if} the observed $\Delta \chi^2_{(\textrm{df})}$ from the data is below some threshold $\chi^2_{\rm (df), thr}$ (which is to be determined empirically based on the adopted $X$ and $P$). 

It is estimated by plotting the distribution of all of our noisy data vectors in the bias$-\Delta\chi_{\rm (df)}^2$ plane, and, for a particular $\Delta \chi^2_{\rm (df), thr}$, evaluating the fraction of points that lie both \emph{above} $\mathrm{bias}=X\sigma$ and \emph{below} $\Delta \chi^2_{\rm (df), thr}$ (i.e., in the lower right quadrant of Figure~\ref{fig:bias_results}).
In practice, one starts by defining the tolerance $X$ and the desired bias confidence $P$. For example, one might require the bias to be smaller than $X=0.3\sigma$ at $90\%$ confidence. Given those numbers, we can then iteratively evaluate Eq.~\eqref{eq:prob_threshold}  with different $\Delta \chi^2_{\rm (df), thr}$ thresholds until the required $P$ is achieved.

\subsection{Dealing with Unphysical $\Delta \chi^2_{(\textrm{df})}$ Values and Unconverged IA Samples}\label{sec:method:unphysical}

It is also worth briefly remarking that in our analysis we found about 50 (out of 1050)
data points for which $\chi^2_{\rm TATT}>\chi^2_{\rm NLA}$, and so $\Delta \chi^2_{(\textrm{df})}<0$. Given that NLA and TATT are nested models, these points are unphysical (a more flexible model should always be able to produce a better or as-good fit). We conclude that they are an artefact of the sampling method; although we tested the robustness of our IS setup, and found it can reproduce the best fit from chains to reasonable precision, some level of sampling noise is still present.
Given this logic, it is reasonable to assume that if we were to find $\Delta \chi^2_{\rm NLA} < \chi^2_{\rm TATT}$ in real data, there would likely be some follow-up investigation and the chains would be rerun. This is particularly true if the $\chi^2$ is an integral part of the analysis plan, as it is in our method.
We thus choose to discard these points. It is worth bearing in mind, however, that this may or may not be a reasonable decision in other setups, depending on the models being compared and the details of the analysis.

Also note that, although our results are based on 21 IA samples, we did initially draw 25 scenarios (see Section~\ref{sec:data:ia_sampling}).
Of these 25, we found four to be so extreme that the NLA \blockfont{PolyChord} chains failed to converge in the noiseless case. These resulted in highly distorted and often bimodal contours in the $S_8-\omegam$ plane, making it difficult to obtain meaningful estimates for the bias. Given this,  and also for the reasoning discussed above, we choose to omit these samples from our analysis. This leaves us with a total of 21 IA samples.

\subsection{The Recommended Method for Model Selection}\label{sec:method:recomendation}

Given all the definitions set out in the sections above, we now follow the recipe outlined below, in order to map and calibrate the bias-metric relation. 
These steps are, in essence, our method; when written out in this form, it can be very easily generalised to other model selection problems beyond our particular  example of IA in cosmic shear.

\begin{enumerate}
    \item \textbf{Sample IA scenarios:} Draw about $10-30$ parameter samples from either a posterior from  a previous analysis or from some reasonable priors using the method described in Section~\ref{sec:data:ia_sampling}  (we used 21 drawn from DES Y1 TATT posteriors).
    
    \item \textbf{Generate data vectors:} Generate a simulated noiseless data vector for each IA scenario drawn in the previous step. Other model parameters (e.g., cosmological and nuisance parameters) should be fixed to some fiducial values. See Section~\ref{sec:data:gen_data}.
    
    \item \textbf{Analyse noiseless data vectors:} There are two chains per data vector, one corresponding to model $M_1$, and the other to $M_2$. Again, we compared the TATT and NLA IA models using \blockfont{PolyChord} to compute statistics. These choices might vary under different applications. Details on the sampling can be found in Section~\ref{sec:method:sampler:nested}.
    
    \item \textbf{Compute parameter bias and plot out the bias-metric relation:} Demonstrate that the noiseless data vectors show a clear correlation between the test statistic (e.g., $\Delta \chi^2_{(\textrm{df})}$ or Bayes factor) and parameter bias, as described in Section~\ref{sec:method:bias}.
    
    \item \textbf{Generate noise realisations:} For each data vector, generate $\sim 50$ noise realisations using the covariance matrix, as explained in Section~\ref{sec:data:noise}. In our case, that gives a total of $21\times50=1050$ noisy data vectors.
    
    \item \textbf{Analyse noisy data vectors:} As discussed above, we choose to use importance sampling to give a fast approximation for the noisy posteriors. For all noise realisations (50) and IA scenarios (21), for both IA models (2) $\longrightarrow$ 2100 total, estimate the posterior, the NLA-TATT bias and the model test statistics. See Section~\ref{sec:modelling:importance_sampling}.   
    
    \item \textbf{Calculate probability:} Plot out the bias-metric relation. Use the quantities computed in the previous step to calculate the probability of bias greater than some pre-defined threshold $X\sigma$ (see Section~\ref{sec:method:bias_probability}).
    
    \item \textbf{Run analyses on blinded data:} Run a full chain on the real data in order to evaluate the observed model statistic. This can then be interpreted in terms of bias probability using the results of the above step. See Section~\ref{sec:results}. 
    
\end{enumerate}

Note that, to obtain an accurate calibration of the $\Delta \chi^2_{(\textrm{df})}$ value, all aspects of the modelling should be as close to the final fiducial analysis setup as possible. For an estimate of the computational resources required to employ the proposed method, see Appendix~\ref{app:computational_resources}.


\section{Results}\label{sec:results}

Now we have outlined the details of our method in Section~\ref{sec:method:recomendation}, we will now consider a specific application. As discussed earlier, we choose to focus on the problem of deciding between two intrinsic alignment models for a cosmic shear analysis: NLA and TATT. 
Although these models \emph{are} nested,  the method does not assume this. 
Indeed, the only requirement is the ability to generate mock data to calibrate the chosen test statistics; therefore it is quite general and can be applied to a variety of model selection problems.

In Section~\ref{sec:results:noiseless} we discuss the results from our \blockfont{PolyChord} chains on noiseless data vectors, and the basic trends. Section~\ref{sec:results:noisy} then discusses the more complete probabilistic calibration, which properly factors in the impact of noise. We also compare our empirical results against theoretically derived $\chi^2$ thresholds. Section \ref{sec:results:nla_fits} looks at how far bias can be inferred from NLA fits alone, without explicit model comparison.
Lastly, Section \ref{sec:results:wider_implications} considers the wider outlook for lensing cosmology.

\subsection{The Noiseless Case}\label{sec:results:noiseless}

\begin{figure}
\includegraphics[width=\columnwidth]{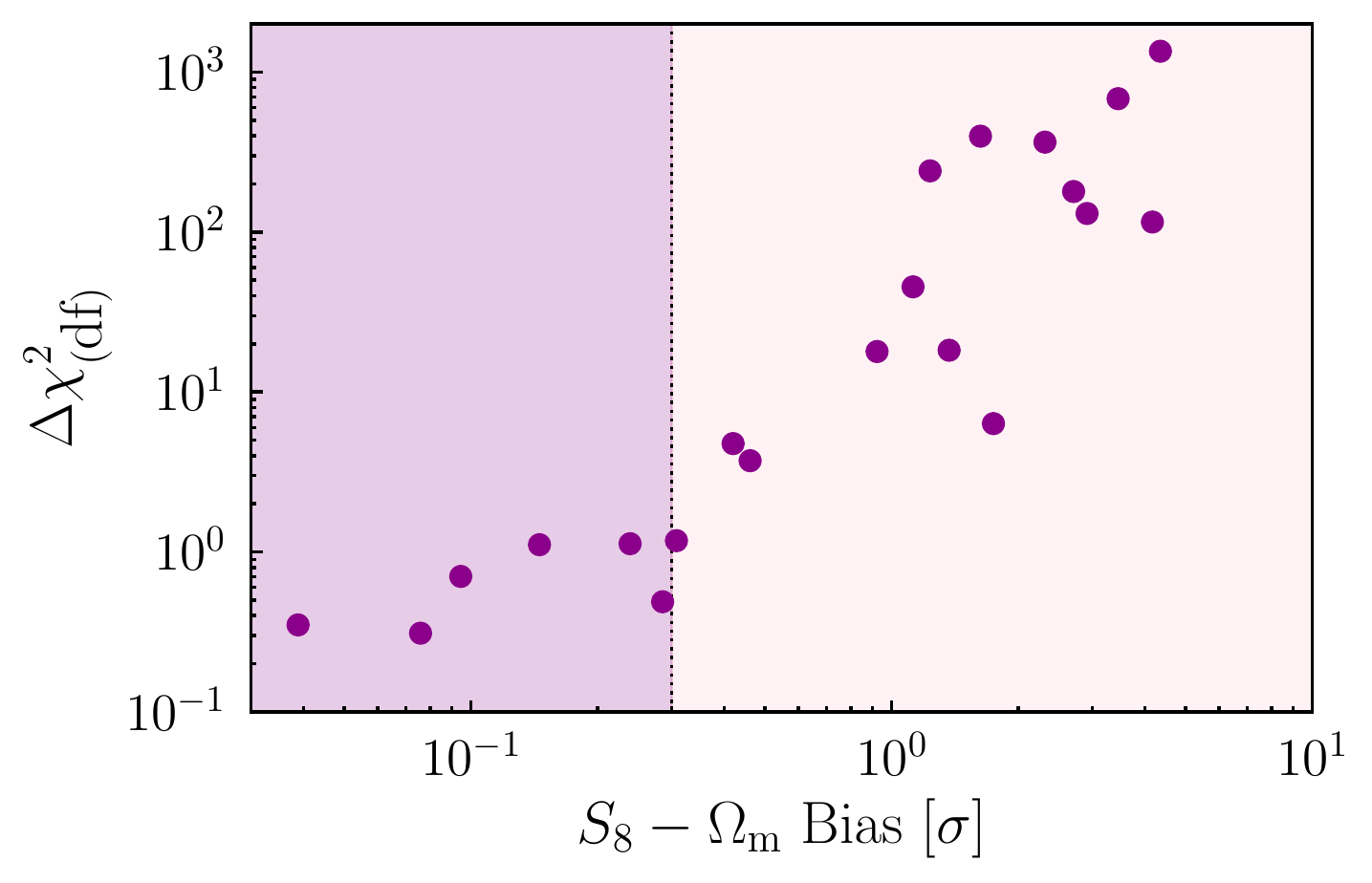}
\caption{
$\Delta \chi^2_{(\textrm{df})}$ as a function of cosmological parameter bias for a DES Y3-like cosmic shear analysis.
The 21 points correspond to noiseless data vectors, generated with different input IA parameters. 
As defined in Eq.~\eqref{eq:deltachi2}, large values of $\Delta \chi^2_{(\textrm{df})}$ indicate that the data prefer TATT over NLA. The vertical dotted line marks the $0.3\sigma$ bias limit used in DES Y3 \citep{krause21}. We see a clear correlation between the observable metric ($\Delta \chi^2_{(\textrm{df})}$) and the underlying parameter bias, particularly for those points for which the bias exceeds $\sim0.2\sigma$.
}
\label{fig:bias_results}
\end{figure}

Considering first the noiseless case, Figure~\ref{fig:bias_results} shows the relation between bias in the $S_8 - \omegam$ plane and the NLA-TATT $\Delta \chi^2_{(\textrm{df})}$.
Each point results from running two chains on the same noiseless simulated data vector, first using NLA, and then using TATT. As defined in Eq.~\eqref{eq:deltachi2}, large values of $\Delta \chi^2_{(\textrm{df})}$ indicate statistical preference for the larger model (i.e., TATT). 
We see a relatively tight relation between bias and $\Delta \chi^2_{(\textrm{df})}$, going from $\Delta \chi^2_{(\textrm{df})} < 1$ when bias is small to relatively large values at the high bias end: 
low bias $\rightarrow$ small $\chi^2$ difference, high bias $\rightarrow$ large $\chi^2$ difference. Interestingly, the relation appears to have the form (approximately) of a double power law, with a steep slope in the high bias regime, switching to a somewhat shallower function below $0.3\sigma$. It is worth stressing, however, that this relation is empirical. We do not have a particular expectation for its shape, and it is likely that the details depend on the analysis choices and survey setup. 
Note that even without data vector noise, this relation presents some scatter. This arises both from sampler noise, and from the fact that this is a complex high dimensional problem, for which two sets of IA values that produce biases of a similar magnitude will not necessarily produce identical $\Delta \chi^2_{(\textrm{df})}$ values.
The vertical dotted line marks the $0.3\sigma$ bias
threshold adopted by DES \citep{krause21}, which we adopt as our fiducial tolerance (see Section~\ref{sec:results:noisy} below). 
Although we cannot use this noiseless result for any empirical method because real data will always contain noise, confirming that these quantities clearly correlate is a necessary first step in our method, and important to check before incurring the expense of further calculations.  We will see in the next section that the correlation between $\Delta\chi^2$ and $S_8-\omegam$ bias holds (with some additional scatter) when we proceed to the noisy case.

In addition to the $\Delta \chi^2_{(\textrm{df})}$, we also consider the Bayes Ratio as a potential model comparison metric; while the former presents a clear relation to the bias (as seen in Figure \ref{fig:bias_results}), we find the latter be a relatively weak indicator, with additional intrinsic scatter. This can be seen in Appendix~\ref{sec:appendixBayes}, and in particular Figure~\ref{fig:bayes_bias_results}.
We also note that $R$ and $\Delta \chi^2_{(\textrm{df})}$ are correlated. In principle one could seek to combine them, but naively treating them as independent metrics is almost certainly double-counting information. 
Therefore, here we focus on the results using $\Delta \chi^2_{(\textrm{df})}$. For further discussion of the Bayes ratio see Appendix~\ref{sec:appendixBayes}.

\subsection{Noise \& Probabilistic Calibration}\label{sec:results:noisy}

\begin{figure}
\includegraphics[width=\columnwidth]{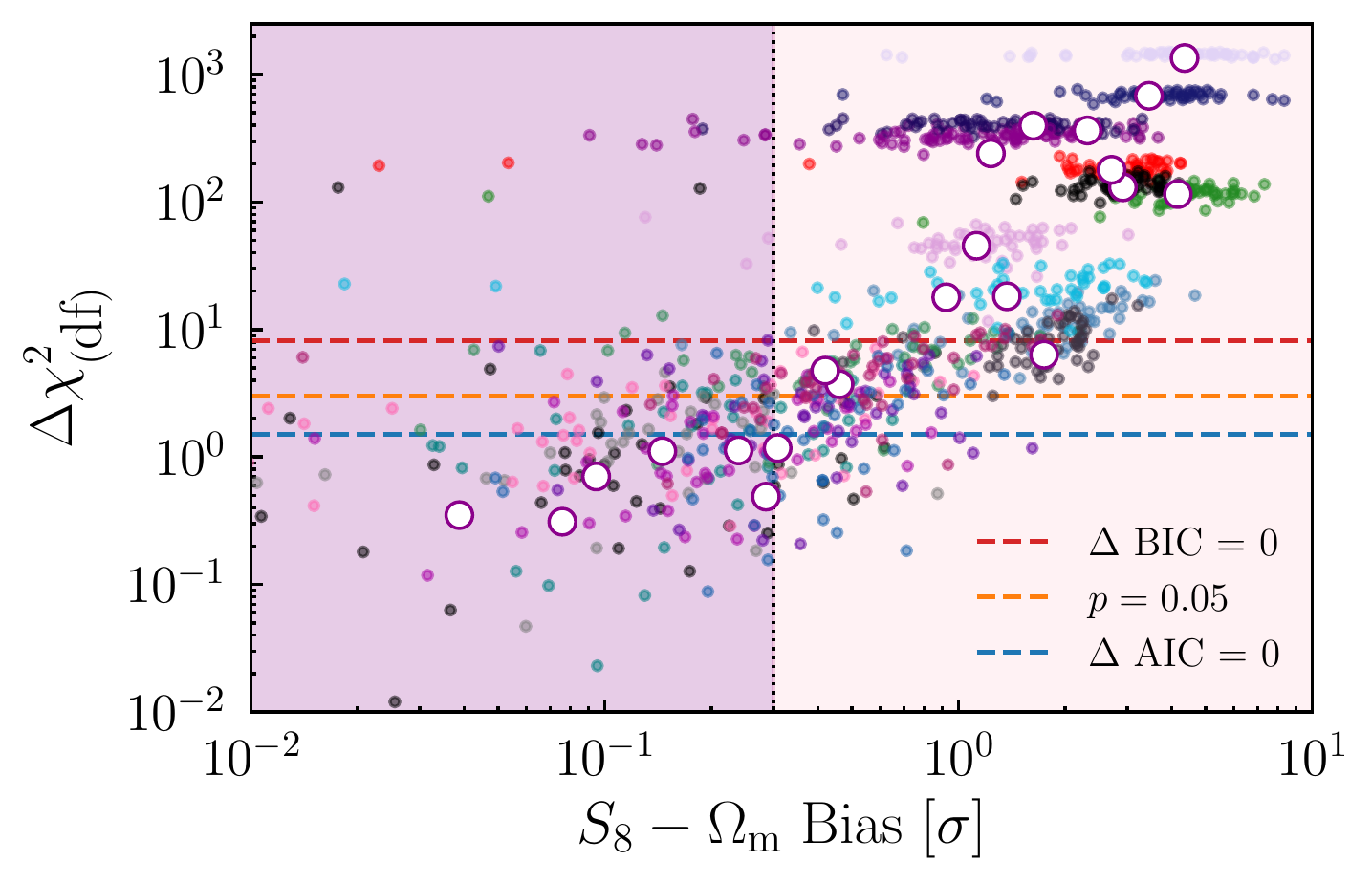}
\caption{The impact of data vector noise on $\Delta \chi^2_{(\textrm{df})}$. 
The larger open points show our 21 IA samples with zero noise (identical to those in Figure \ref{fig:bias_results}). The smaller coloured dots show the effect of adding random noise realisations, for which parameter constraints are estimated using importance sampling. For each of the 21 colours, we have a collection of 50 realisations.
The red and blue horizontal dashed lines mark threshold $\Delta \chi^2_{(\textrm{df})}$ values, defined by the points where the BIC and AIC respectively prefer NLA and TATT equally. The orange dashed line corresponds to a $p-$value $p(\Delta \chi^2)=0.05$ (see text, Section~\ref{sec:results:noisy}).  The fact that these formal cut-offs are relatively ineffective in isolating the bias $<0.3\sigma$ region motivates us to adopt an empirical approach.
}
\label{fig:noise_bias_results}
\end{figure}

In Figure \ref{fig:noise_bias_results} we illustrate the impact of data vector noise in the bias-$\Delta \chi^2_{(\textrm{df})}$ plane. We show the same 21 noiseless samples discussed above, but now overlain with multiple different noise realisations, as approximated using importance sampling. As we can see, noise introduces scatter in the bias-$\Delta \chi^2_{(\textrm{df})}$
relation.  While this noise is considerably less than in the case of the Bayes factor   (for which we show in Appendix~\ref{sec:appendixBayes} that the scatter due to noise is so large that the relation with bias is extremely weak), it is still non-negligible.

For comparison, we show the $\Delta \chi^2_{(\textrm{df})}$ cut-offs implied by some standard model selection metrics: BIC, AIC and a $p-$value significance threshold\footnote{Assuming that the TATT model has 2 additional effective degrees of freedom compared with NLA. Note that this $df_\text{diff}$ value was calculated for the DES Y3 shear-only (no shear ratio) case in \citet*{y3-cosmicshear2}. It is thus valid for our particular case, but would not necessarily hold under changes to the data vector or analysis choices.} of $p=0.05$
(see Section~\ref{sec:method:stats:chi2} for definitions). 
Unfortunately, in the presence of noise, we see that all three cut-offs are relatively weak indicators of bias -- i.e., they still favour the simpler model even when significant amount of bias has been introduced in the cosmological parameters. Even in the case of AIC, which is the strictest of the three, there are a problematic fraction of noise realisations where the observable metric favours NLA, and yet NLA is biased by $>0.3\sigma$
(see the points in the lower right hand corner of Figure \ref{fig:noise_bias_results}). 
This illustrates a key motivation for adopting an empirical calibration. Theoretical limits imposed using, e.g., $p-$values are not designed to optimise the quantities we care most about (i.e., parameter biases). For a given analysis, it is impossible to know from first principles what level of bias is excluded for a given statistical metric cut-off without some form of calibration.

These observations have important consequences for our method. Since the exact noise realisation in any real data set is unknown, one cannot simply run a single set of IA samples (as in Figure~\ref{fig:bias_results}), and perform a 1:1 bias-$\Delta \chi^2_{(\textrm{df})}$ mapping. Nor, as we can see from Figure~\ref{fig:noise_bias_results}, can we simply fall back on theoretical cut-offs to reliably guard against model bias. Instead, we must consider the problem as a probabilistic one, and factor in the uncertainty from noise.

\subsubsection{Probabilistic Interpretation}

\begin{figure}
\includegraphics[width=\columnwidth]{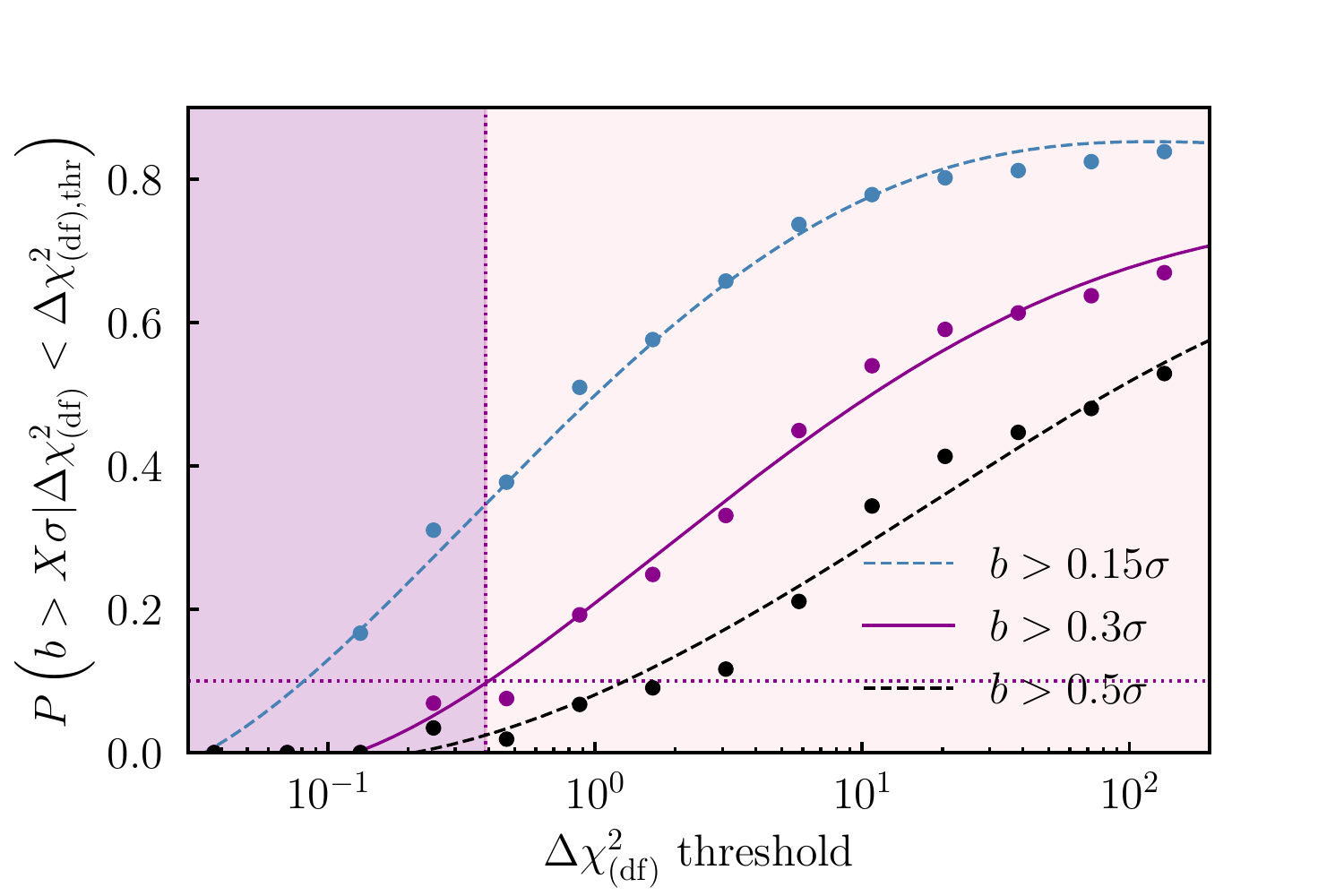}
\caption{Probability of exceeding some specified level of cosmological parameter bias, as a function of the threshold $\Delta \chi^2_{(\textrm{df})}$ value. 
For a DES Y3-like cosmic shear data vector with unknown noise and IA realisation, and that is found to give an observed $\Delta \chi^2_{(\textrm{df})}$ lower than threshold the $\Delta \chi_{\rm (df) thr}^2$, $P$ is defined as the probability that the results using NLA are biased by more $X \sigma$ in the $S_8-\Omega_{\rm m}$ plane. Different values of $X$ are represented by different colours. In each case, we  show both the direct measurement of $P$ using importance sampling (coloured points), and the lines are obtained by doing a polynomial fit.
For illustrative purposes, we also show the $\Delta \chi^2_{(\textrm{df})}$ threshold that would guarantee NLA is unbiased to within $0.3\sigma$ at a confidence level of $90\%$ (dotted lines and shading).}
\label{fig:bias_prob_chi2_threshold}
\end{figure}

To interpret our results in a quantitative way, we use Eq.~\eqref{eq:prob_threshold} and calculate the bias probability $P(b>X\sigma | \Delta \chi^2_{\rm (df), obs}< \Delta \chi^2_{\rm (df), thr})$. This quantity should be interpreted as the conditional likelihood that, \emph{if} in the real data one finds a $\Delta \chi^2_{\rm (df), obs}$ value below some limit $\Delta \chi^2_{\rm (df), thr}$ (a horizontal line in Figure~\ref{fig:noise_bias_results}), then the analysis using NLA will still in fact be biased by $X\sigma$ or more.

Figure~\ref{fig:bias_prob_chi2_threshold} shows three curves corresponding to chosen bias thresholds of $0.15\sigma$, $0.3\sigma$, and $0.5\sigma$. Each point is calculated using Eq.~\eqref{eq:prob_threshold} and the curves are obtained by fitting a fifth order polynomial to the points. We tested the stability of these smoothed fits, and found that they are robust to doubling the number of noise realisations in Figure~\ref{fig:noise_bias_results} (from 50 per IA sample to 100). 
This result provides a powerful tool, which can be used to interpret results from real data. For instance, say we were to run NLA and TATT chains on a blinded Y3 data vector, and find $\Delta \chi^2_{\rm (df), obs} < 0.4$. With the aid of Figure~\ref{fig:bias_prob_chi2_threshold}, we could say that the chance of the NLA run being biased by more than $0.5\sigma$ in $S_8-\omegam$ is about $3\%$. The probability of exceeding a $0.3\sigma$ threshold is about $10\%$, and the chance of bias greater than $0.15\sigma$ is about $37\%$.
In practice, the bias tolerance is an analysis choice. As discussed previously, DES Y3 chose a value of $0.3\sigma$ by which to judge simulated chains. The exact number, however, is somewhat subjective, and the most convenient value may depend on how well sampled the low bias end of the bias-$\Delta \chi^2_{(\textrm{df})}$ relation is.
As one might expect, the lower the bias threshold $X$, the stronger the requirement on the $\Delta \chi^2_{(\textrm{df})}$ (i.e., the stronger the data needs to favour NLA) in order to keep the bias probability $P(b>X\sigma)$ low.

To understand how our results depend on various analysis choices, it is perhaps useful to think of the process in Section~\ref{sec:method:recomendation} as a series of transformations between different distributions. The points in Figure~\ref{fig:noise_bias_results}, which determine the final $\chi^2$ threshold, can be thought of as the convolution of two parts: an initial distribution of noiseless points $P_s(\Delta \chi^2_{(\textrm{df})}, b)$ (the open points in Figure~\ref{fig:noise_bias_results} and the filled in Figure~\ref{fig:bias_results}) and a second distribution conditioned on each one $P_N(\widetilde{\Delta \chi^2_{(\textrm{df})}}, \tilde{b} | \Delta \chi^2_{(\textrm{df})}, b)$ (where the tilde denotes noisy values of $\Delta \chi^2_{(\textrm{df})}$ and bias). In the first case, $P_s$, we start with a distribution in IA parameter space $P(\boldsymbol{\theta}_{\rm IA})$, which we choose. The samples from $P(\boldsymbol{\theta}_{\rm IA})$ are mapped onto a distribution of noiseless data vectors, which are then transformed (via running chains) into samples in the final bias-$\Delta \chi^2_{(\textrm{df})}$ space: $P(\boldsymbol{\theta}_{\rm IA}) \rightarrow P(\boldsymbol{D}) \rightarrow P_s(\Delta \chi^2_{(\textrm{df})}, b)$. Both mapping steps are dependent on the survey analysis choices (choice of power spectrum, $n(z)$, covariance matrix, etc.). This is not a problem, as long as these choices match the ones that will be applied on real data. It is, however, likely these choices have an impact on the observed bias-$\Delta \chi^2_{(\textrm{df})}$ correlation.
It is clear from this that $P_s(\Delta \chi^2_{(\textrm{df})}, b)$ also depends to an extent on the choice of $P(\boldsymbol{\theta}_{\rm IA})$. We can see that $P(\boldsymbol{\theta}_{\rm IA})$ behaves analogously to a prior, restricting the range of possibilities in the subsequent steps. However, given that the purple points in Figure~\ref{fig:bias_results} show a relatively tight correlation and cover a broad range of bias relatively uniformly, we do not expect the details to change things considerably.

The other part of the final sampling of points is the noise distribution $P_N(\widetilde{\Delta \chi^2_{(\textrm{df})}}, \tilde{b} | \Delta \chi^2_{(\textrm{df})}, b)$. We obtain this for a particular IA sample by sampling noise realisations, and so transforming $P(\widetilde{\boldsymbol{D}}|\boldsymbol{D}) \rightarrow P_N(\widetilde{\Delta \chi^2_{(\textrm{df})}}, \tilde{b} | \Delta \chi^2_{(\textrm{df})}, b)$. This process is again dependent on the covariance matrix, but not on the choice of $P(\boldsymbol{\theta}_{\rm IA})$ (at least, not directly). 

The end result of the above is that, by convolving to get to $P(\widetilde{\Delta \chi^2_{\rm (df)}}, \tilde{b})$, we are able to map out the relationship between a quantity we can measure (the noisy $\widetilde{\Delta \chi^2_{\rm (df)}}$) and the one we are interested in (parameter bias $\tilde{b}$).

\subsubsection{Bias Tolerance Implications}

\begin{figure}
\includegraphics[width=\columnwidth]{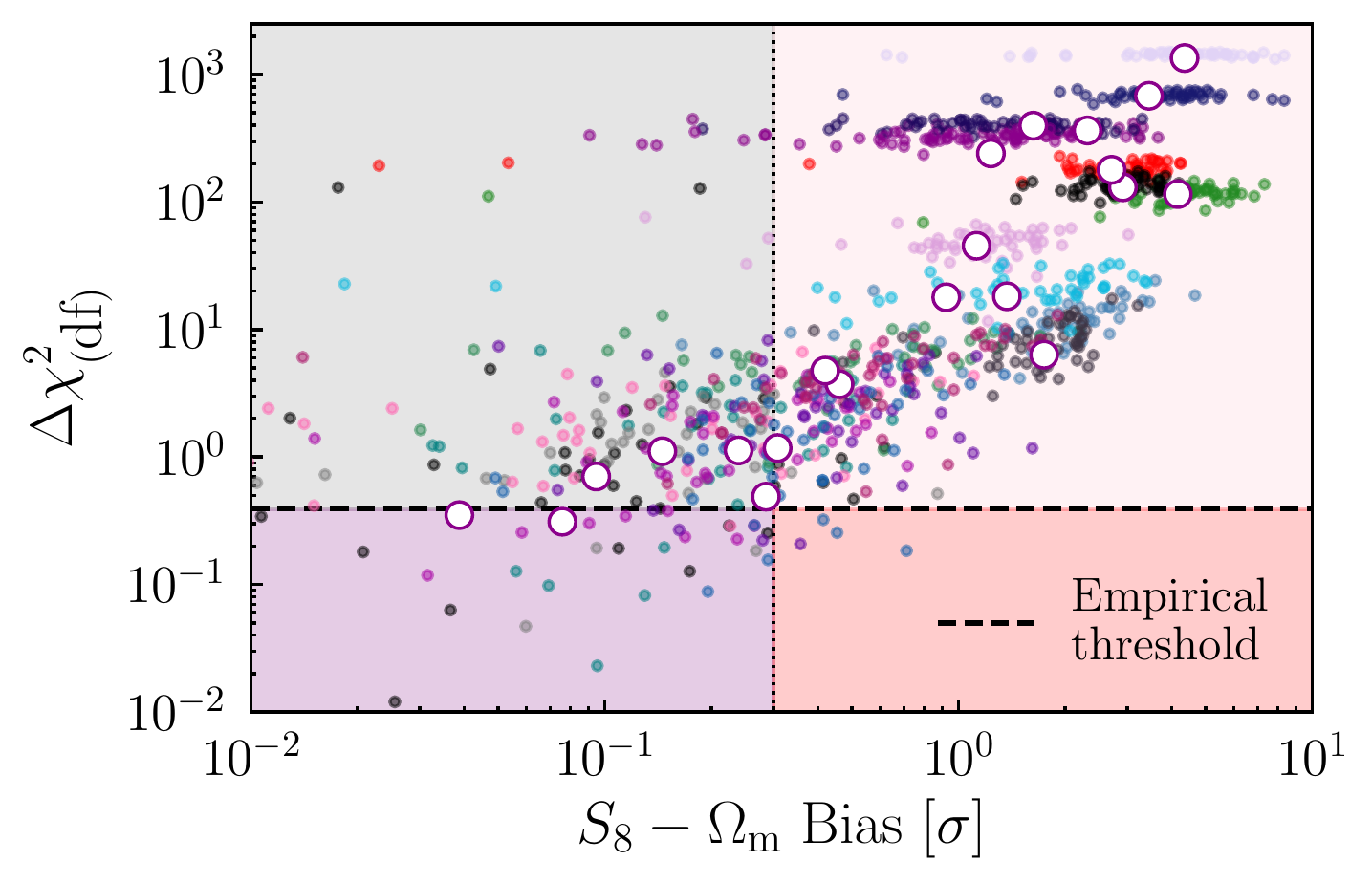}
\caption{The impact of data vector noise on $\Delta \chi^2_{(\textrm{df})}$. The points are the same as in Figure~\ref{fig:noise_bias_results}. 
The horizontal line represents an empirical $\Delta \chi^2_{(\textrm{df})}$ threshold, derived to ensure bias below $0.3\sigma$ with $90\%$ confidence. 
The four different shaded regions distinguish the following possible scenarios: purple - NLA is sufficient and the calibrated $\Delta \chi^2_{(\textrm{df})}$ favours NLA ; grey - NLA would be sufficient and yet $\Delta \chi^2_{(\textrm{df})}$ chooses TATT ; pink - NLA is insufficient and $\Delta \chi^2_{(\textrm{df})}$ favours TATT ; Red - NLA is insufficient and yet $\Delta \chi^2_{(\textrm{df})}$ still chooses NLA. This last case is the most dangerous, and the $\Delta \chi^2_{(\textrm{df})}$ threshold is chosen to keep the fraction of points in this quadrant acceptably small.
}
\label{fig:noise_bias_results_regions}
\end{figure}

We further illustrate our results by taking a concrete example. For our DES Y3 setup, we choose a bias tolerance of $X=0.3\sigma$, and a bias probability of $10\%$ ($P=0.1$). Using Figure~\ref{fig:bias_prob_chi2_threshold}, this gives us $\Delta \chi^2_{\rm (df), thr}=0.4$ (reading across where the horizontal dashed line meets the purple curve), which is shown in Figure~\ref{fig:noise_bias_results_regions} (the horizontal line labelled ``empirical threshold").
With the bias and $\Delta \chi^2_{(\textrm{df})}$ thresholds fixed, the four shaded regions in Figure \ref{fig:noise_bias_results_regions} distinguish the following possible scenarios: (a) NLA is sufficient (i.e., the bias is below our $0.3\sigma$ limit) and $\Delta \chi^2_{(\textrm{df})}$ chooses NLA (i.e., $\Delta \chi^2_{(\textrm{df})}<\Delta \chi^2_{\rm (df), thr}$; purple); (b) NLA is sufficient and $\Delta \chi^2_{(\textrm{df})}$ chooses TATT (grey); (c) NLA is insufficient and $\Delta \chi^2_{(\textrm{df})}$ chooses TATT (pink); (d) NLA is insufficient and $\Delta \chi^2_{(\textrm{df})}$ chooses NLA (red). As we discussed previously, case (d) is the most dangerous, for obvious reasons. Scenario (b) is not ideal (since we may end up with a model that is more complicated than strictly necessary), but does not result in cosmological parameter biases.
The different scenarios can be better understood with the help of a confusion matrix, shown in Table~\ref{tab:confusion_matrix}.

\begin{table}
    \centering
    \caption{Confusion Matrix. The samples are split into quadrants, corresponding to the four shaded regions in Figure \ref{fig:noise_bias_results_regions}. 
    The left/right columns show the fraction of IA samples that give a bias above and below $0.3\sigma$. The rows indicate whether or not our method using the calibrated $\Delta \chi^2_{(\textrm{df})}$ prefers NLA or TATT.}

    \begin{tabular}{l|l|c|c|c}
        \multicolumn{1}{l}{}&\multicolumn{3}{c}{Model preferred by Bias} & \\
        \cline{3-4}
        \multicolumn{2}{c|}{}& NLA & TATT &\multicolumn{1}{c}{Total}\\
        \cline{2-4}
        \multirow{2}{*}{Model preferred} & TATT & $19.1\%$ & $77.1\%$ & $96.2\%$\\
        \cline{2-4}
        \multirow{2}{*}{by $\Delta \chi^2_{(\textrm{df})}$} & NLA & $3.4\%$ & $0.4\%$ & $3.8\%$\\
        \cline{2-4}
        \multicolumn{1}{c}{} & \multicolumn{1}{c}{Total} & \multicolumn{1}{c}{$22.5\%$} & \multicolumn{    1}{c}{$77.5\%$} & \multicolumn{1}{c}{$100\%$}\\
    \end{tabular}
    \label{tab:confusion_matrix}
\end{table}

\noindent
The columns here represent the model preference according to the amount of bias, and the rows represent the model preference according to the $\Delta \chi^2_{(\textrm{df})}$. 
Since we are effectively using $\Delta \chi^2_{(\textrm{df})}$ as an empirical proxy for bias, we treat the classification according to the latter (i.e., does using NLA cause cosmological parameter biases for a particular data vector exceeding $0.3 \sigma$?) as the truth and the label according to the former (i.e., is $\Delta \chi^2_{(\textrm{df})}$ below $\Delta \chi^2_{\rm (df), thr}$?) as the prediction (in machine learning language).

We see that in our samples, the cosmological parameter bias indicates that NLA should be preferred about $22.5\%$ of the time, while TATT should be preferred $77.5\%$. In other words, NLA introduces a bias smaller than our threshold in $\sim 1/4$ of the cases.
Note that this fraction is somewhat dependent on our particular choices. A different choice of posteriors in Figure~\ref{fig:y1_contours}, for example, could change this fraction. We do not, however, expect this to affect the validity of the method.

When it comes to the performance of $\Delta \chi^2_{(\textrm{df})}$ in identifying the correct model, we see that it favours TATT in $96.2\%$ of the cases, and NLA in only $3.8\%$. We can see that our method is quite conservative, in the sense that there is a non-negligible false positive rate. That is, it prefers TATT over NLA in $19.1\%$ of cases, even though NLA would not introduce bias to the model above the $0.3\sigma$ threshold. 
Reassuringly, however, we also see that our method is is highly effective in ruling out real bias. The strongest feature 
of our approach,
perhaps, is the fact that it is very unlikely to select NLA if it is, in fact, introducing biases to the analysis. We can see this by the very small population of points in the lower right of the matrix (and the red shaded quadrant in Figure~\ref{fig:noise_bias_results_regions}): this happens in only $\sim 0.4\%$ of cases. 
Put another way, \emph{if} the calibrated $\Delta \chi^2_{(\textrm{df})}$ favours TATT, there is a roughly $20\%$ chance ($19.1/96.2$) that NLA would, in reality, have been fine. Conversely, \emph{if} it prefers NLA, there is $\sim10\%$ ($0.4/3.8$) that NLA is insufficient.
Therefore, even though the end result is somewhat cautious (in that there is a moderate false positive rate for TATT), on the positive side we can be confident that if NLA is in fact preferred by the data, it is very unlikely that it will introduce biases to the analysis.
As a remark, however, it is important to acknowledge that a possible conclusion from these results is that simply using the most general model is the cheapest alternative from the perspective of computational resources.  It is not obvious that this will always be the case, however, given the dependence on analysis setup and other factors.

It is also worth noticing that although the above discussion applied for our specific choices, we can control the conservatism to a significant degree through our analysis choices. We chose a specific bias tolerance and probability that we considered realistic. By changing these values (for example, allowing a bias probability of $20\%$, or $25\%$) one can effectively shift the position of the cross in Figure~\ref{fig:noise_bias_results_regions}, and trade off false positives for false negatives. This is another advantage of the method: it makes the level of conservatism explicit (and indeed quantifiable), and allows one to adjust that level as preferred. This is much less true when using alternative approaches to model selection.

\subsection{A Simpler Approach: How Much Can We Tell From A Single Model?}\label{sec:results:nla_fits}

It is also worth taking a moment to consider a related question: if the true IA scenario is extreme enough to give significant cosmological biases, would there be clear red flags from NLA alone, assuming that no fits were carried out with TATT? If this were the case, it would provide a simpler route -- instead of performing empirical calibration using simulations, we could simply run one model on the data, and interpret results to see if a more sophisticated model is needed.  
Considering our 21 IA scenarios with a fixed noise realisation, we find that around $50\%$ of cases with bias $>0.3\sigma$ have $\chi^2$ values that appear entirely consistent with being drawn from the corresponding $\chi^2$ distribution\footnote{Where the NLA $p-$value here is calculated by assuming $\chi^2_{\rm NLA}$ is drawn from a $\chi^2$ distribution with 224 degrees of freedom (see \citealt*{y3-cosmicshear2}). The null hypothesis in this case is that the NLA model is adequate to describe the data, and so small $p-$values would indicate model insufficiency.}, $p(\chi^2_{\rm NLA})>0.05$. A similar picture is seen when we consider a single high bias IA scenario with alternative noise realisations -- computing $p-$values for each realisation, the majority are above 0.05, even in the presence of bias $>0.3\sigma$. 
In other words, even in cases where NLA is significantly biased, it is not necessarily obvious from considering the uncalibrated value of $\chi^2_{\rm NLA}$ alone. In contrast, the method we propose, using $\Delta \chi^2_{(\textrm{df})}$, can correctly identify the need to use TATT to achieve sufficiently unbiased results $77.1/77.5\sim99\%$ of the time (see the confusion matrix in Section~\ref{sec:results:noisy}). 

\begin{figure}
\includegraphics[width=\columnwidth]{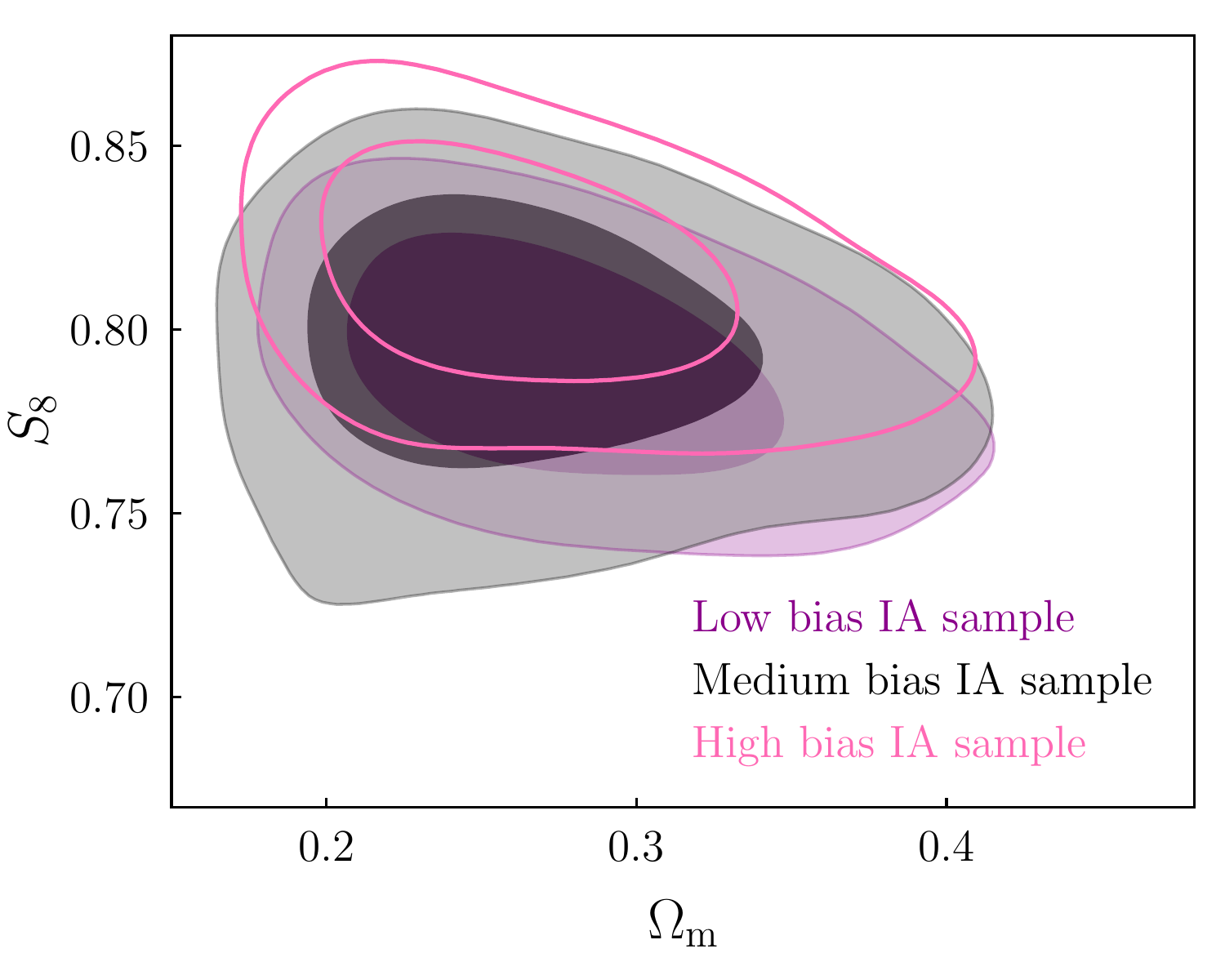}
\includegraphics[width=\columnwidth]{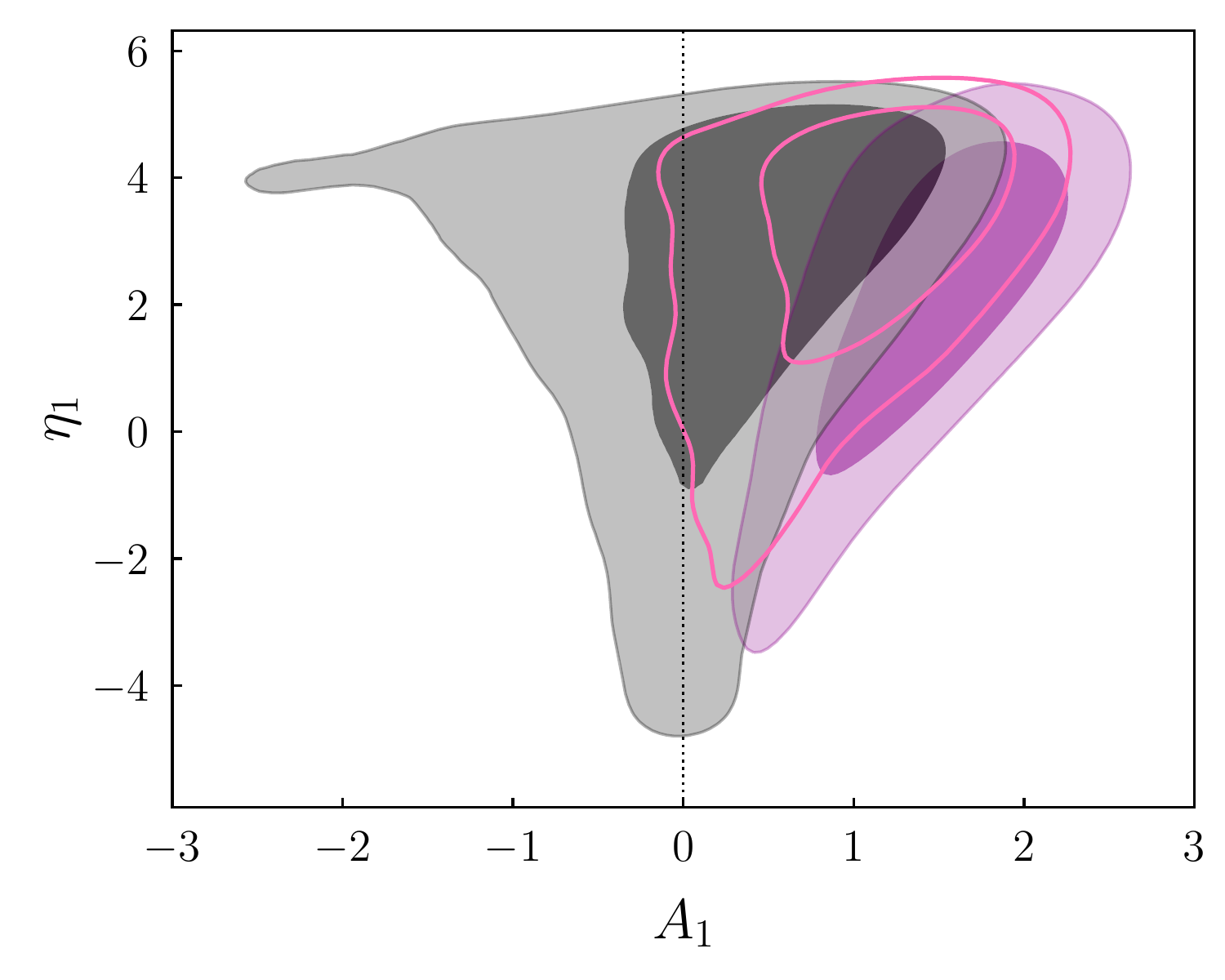}
\caption{Examples of the simulated NLA posteriors from three particular IA scenarios with our fiducial noise realisation. These samples were chosen to span a range of bias levels (as defined relative to the TATT posteriors from the same data vectors). In order of severity, the low bias case (purple) has a bias in the $S_8-\omegam$ plane of $\sim0.1\sigma$, and $\Delta \chi^2_{(\textrm{df})}=0.24$, $R=21.9\pm6.1$; the medium bias case (black) has $0.36\sigma$ bias, $\Delta \chi^2_{(\textrm{df})}=0.49$, $R=1.5\pm0.3$; the high bias case (pink, open) has $0.82\sigma$ bias, $\Delta \chi^2_{(\textrm{df})}=1.98$, $R=1.1\pm0.2$. In all cases the posteriors are not visibly distorted (although in the medium and high bias cases, the $\eta_1$ posterior is cut off slightly by the upper prior edge at $\eta_1=5$).
}
\label{fig:nla_contours}
\end{figure}

Likewise, although extreme biases do tend to distort the shape of the posteriors, this is not always true in more moderate (but still significantly biased) cases. Figure~\ref{fig:nla_contours} shows the NLA posteriors in a few different IA scenarios, spanning the range from almost no bias (purple shaded), to $\sim 1 \sigma$ (pink, open contours). For reference we also show the posteriors from TATT fits to the same data vectors in Appendix~\ref{app:posteriors}. Taken in isolation, none of these show clear signs of problems with the model. It is also interesting that IA mismodelling bias does not always translate into significantly non-zero values for the inferred NLA parameters. In the medium bias case, for example, $A_1$ and $\eta_1$ are both consistent with zero to $<1\sigma$. 
Here there is a relatively strong degeneracy between $A_1$ and $S_8$, allowing both $A_1\sim0$ combined with low $S_8$, but also a stronger IA amplitude $(A_1\sim1)$ with a larger $S_8$. In projection, this results in broad contours on both parameters (notice the black contours in the upper panel of Figure~\ref{fig:nla_contours} are slightly wider than the others, with a more prominent asymmetry at high/low $S_8$).

\subsection{Intrinsic alignment modelling \& wider implications for weak lensing}\label{sec:results:wider_implications}

The results discussed so far have a number of direct implications for the question of intrinsic alignment model selection. Primarily, we have shown that it is \emph{possible} to perform empirical model selection with lensing data. There is a clear relation between cosmological parameter bias and $\Delta \chi^2_{(\textrm{df})}$, 
which allows one to define a threshold that can then be applied to the real data. 
That said, the failure of conventional statistical metrics (e.g., $p-$values) to identify scenarios with significant cosmological parameter biases is notable, and should be kept in mind when trying to understand statistics derived from any single run on real data.
The properly calibrated model statistics, however, provide an alternative to the model selection exercises used in previous analyses, which have tended to rely on either simulated analyses (\citealt*{y3-cosmicshear2} Section A3), or arguments based on direct-detection studies (\citealt{hikage19,joachimi21}, Sections 5.4 and 2.4 respectively). The empirical method is arguably an advance on both; first of all, it avoids questions about what constitutes an ``extreme" model, which tend to arise in the simulation-based approach. Since the current best constraints on TATT model parameters are relatively weak, it is relatively easy to select an IA scenario that is both consistent with observations, and which would cause significant bias in an NLA analysis (note that this is still true in light of the most recent DES Y3 results \citealt*{y3-cosmicshear2}; \citealt{y3-cosmicshear1,y3-kp}). 
Our empirical approach also avoids the uncertainties that are inherent in extrapolating observations based on direct IA measurements on one very specific type of galaxies to weak lensing measurements on another population entirely.

Although the empirical approach has various strengths, is worth reiterating that data vector noise is a significant source of scatter in the bias-$\Delta \chi^2_{(\textrm{df})}$ relation. For this reason it is important to accurately simulate the noise properties of the particular data set. While this is in principle simple, given an accurate covariance matrix, it does mean the model selection exercise needs to be repeated for any new data set or changes to the analysis. It also means it is crucial to have a fast and accurate way of estimating posteriors for a large number of noise realisations, such as the IS framework used here.

It is also interesting, finally, to consider how our findings relate to the real Y3 results. Although comparing with the full $3\times2$pt results is difficult, for the reasons given above, \citet*{y3-cosmicshear2} (Section VIIB and Table III)
present a comparison of IA models without shear ratios, an analysis configuration that matches ours. Specifically, 
comparing the 2 parameter NLA model with 5 parameter TATT, they 
find $R=1.70 \pm 0.36$ and $\Delta \chi^2_{(\textrm{df})}=5.2/2=2.6$. Interpreted with the help of Figure \ref{fig:bias_prob_chi2_threshold}, this puts the risk of NLA being biased by $>0.3\sigma$ at somewhere around $30\%$, meaning runs using NLA on Y3 were more likely than not to be unbiased to within the $0.3\sigma$ threshold.

\section{Conclusions}\label{sec:conclusions}

In this paper, we explore the idea that model selection for cosmological analyses could be performed \emph{a posteriori}, being informed by the blinded data themselves. Our goal is to select a model that is sufficient to describe the data, resulting in unbiased parameter constraints at some specified tolerance level. We chose to focus on a specific problem: how best to decide on an intrinsic alignment model for a cosmic shear analysis. This is an important question, and one that has been the subject of much discussion within the weak lensing community in recent years. 
That said, the basic concept behind our method is much more general, and could be applied in a variety of different contexts; in principle it requires only that the data (including its noise) can be readily simulated.

Using simulated noisy DES Y3 weak lensing data, we tested the method, and identified statistical tools with which to implement it.
The main conclusions of our study are as follows:
\begin{itemize}
\item We showed a clear relation between the $\chi^2$ difference between two models, and model insufficiency bias on cosmological parameters. This relation was seen to extend across a wide range of biases, from low to high, allowing one to define an empirical $\Delta \chi^2$ threshold in order to ensure bias is below an acceptable level. 
\item We tested a number of common $\chi^2$-based metrics such as AIC, BIC and pre-defined $p-$value cutoffs. These were seen to be generally under-cautious, favouring the simpler model even in the presence of $1-2\sigma$ parameter biases. This result motivates us to use an empirical $\Delta\chi^2$ calibration. Similarly, when trying to interpret the goodness of fit statistic from a single model, a standard $p(\chi^2)=0.05$ cutoff is not reliable to rule out significant biases.
\item In addition to maximum likelihood-based statistics, we also consider the Bayes factor as a model selection tool. Although useful in the extreme cases, it was seen to be only weakly discriminating for cosmological parameter biases in the range $0.2-1\sigma$. 
We therefore recommend it be used with caution, ideally in conjunction with other model selection metrics.
\item Noise is seen to have a potentially significant impact on both the cosmological bias, and the $\Delta \chi^2_{(\textrm{df})}$, for any given input IA scenario. At high bias, the picture is relatively stable; noise cannot, in general, cause model selection metrics to prefer the simpler model in a case where adopting that model induces large cosmological parameter biases. The reverse is, however, possible. Due to noise, one can end up in scenario with small cosmological parameter bias, but with selection metrics favouring the more complex model. In this regard, our method tends to err on the side of caution.
\end{itemize}

\noindent
Although our qualitative findings are general, it is worth bearing in mind that the details are specific to the DES Y3 cosmic shear only setup. Factors such as choice of two-point statistics, covariance and scale cuts could very easily have an impact, as could modelling choices (baryonic treatment, power spectrum, cosmological model etc) and the choice of sampler. 
It is therefore important that the simulated analyses used to derive a $\Delta\chi^2$ threshold are as close as possible (and ideally identical) to the real setup that will be applied to the blinded data.  

Model selection is an important topic in cosmology, and in science more generally. It is quite common to have a set of models 
under consideration, with little prior knowledge about the values of their parameters; what level of complexity is sufficient to describe the data, given its precision, depends on the unknown true model and its unknown parameter values.
Given these circumstances, arguably the most cautious approach would be to use the most flexible model, which is more likely to be unbiased.  
This paper sets out an alternative method, which allows information in the data to inform model selection. Although applicable in similar situations to Bayesian Model Averaging \citep[BAM;][]{liddle06b,vardanyan11}, i.e., where there is not enough prior information to justify choosing one model over another,
our approach has the advantage of simplicity, and maintains the idea of a fiducial model, which is often useful for practical purposes.
It also avoids the prior dependence of methods such as BAM, which is well documented in the literature. Given its generality, simplicity, and the relatively low level of resources required, we foresee applications of the empirical method discussed in this paper to future analyses as a model selection tool in many contexts.

\section*{Data Availability}

All simulated data vectors, \blockfont{PolyChord} chains and Importance Sampling noise samples used in this work are publicly available at \url{https://github.com/AndresaCampos/empirical_model_selection}.

\section*{Acknowledgements}

We thank Scott Dodelson, Sukhdeep Singh, Lucas Secco, Alex Amon, Judit Prat and Agn\`es Fert\'e  for useful discussions contributing to this work. Many thanks also to Jessie Muir, Noah Weaverdyck, Ot\'avio Alves, Shivam Pandey, and Cyrille Doux for support with code, in particular with setting up the importance sampling pipeline used in this paper. Contour plots were made using the \blockfont{GetDist} package \citep{lewis19}.

Andresa Campos thanks the support from the U.S. Department of Energy grant DE-SC0010118 and the NSF AI Institute: Physics of the Future, NSF PHY-2020295. Simon Samuroff is partially supported by NSF grant AST-2206563. RM is supported in part by the Department of Energy grant DE-SC0010118 and in part by a grant from the Simons Foundation (Simons Investigator in Astrophysics, Award ID 620789).

\bibliographystyle{mnras}
\bibliography{refs}

\begin{thebibliography}{}
\makeatletter
\relax
\def\mn@urlcharsother{\let\do\@makeother \do\$\do\&\do\#\do\^\do\_\do\%\do\~}
\def\mn@doi{\begingroup\mn@urlcharsother \@ifnextchar [ {\mn@doi@}
  {\mn@doi@[]}}
\def\mn@doi@[#1]#2{\def\@tempa{#1}\ifx\@tempa\@empty \href
  {http://dx.doi.org/#2} {doi:#2}\else \href {http://dx.doi.org/#2} {#1}\fi
  \endgroup}
\def\mn@eprint#1#2{\mn@eprint@#1:#2::\@nil}
\def\mn@eprint@arXiv#1{\href {http://arxiv.org/abs/#1} {{\tt arXiv:#1}}}
\def\mn@eprint@dblp#1{\href {http://dblp.uni-trier.de/rec/bibtex/#1.xml}
  {dblp:#1}}
\def\mn@eprint@#1:#2:#3:#4\@nil{\def\@tempa {#1}\def\@tempb {#2}\def\@tempc
  {#3}\ifx \@tempc \@empty \let \@tempc \@tempb \let \@tempb \@tempa \fi \ifx
  \@tempb \@empty \def\@tempb {arXiv}\fi \@ifundefined
  {mn@eprint@\@tempb}{\@tempb:\@tempc}{\expandafter \expandafter \csname
  mn@eprint@\@tempb\endcsname \expandafter{\@tempc}}}

\bibitem[\protect\citeauthoryear{{Akaike}}{{Akaike}}{1973}]{akaike73}
{Akaike} H.,  1973, Proceedings of the Second International Symposium on
  Information Theory, Tsahkadsor, Armenia, USSR

\bibitem[\protect\citeauthoryear{{Amon} et~al.,}{{Amon}
  et~al.}{2022}]{y3-cosmicshear1}
{Amon} A.,  et~al., 2022, \mn@doi [\prd] {10.1103/PhysRevD.105.023514}, \href
  {https://ui.adsabs.harvard.edu/abs/2022PhRvD.105b3514A} {105, 023514}

\bibitem[\protect\citeauthoryear{{Andrae}, {Schulze-Hartung}  \&
  {Melchior}}{{Andrae} et~al.}{2010}]{andrae10}
{Andrae} R.,  {Schulze-Hartung} T.,   {Melchior} P.,  2010, arXiv e-prints,
  \href {https://ui.adsabs.harvard.edu/abs/2010arXiv1012.3754A} {p.
  arXiv:1012.3754}

\bibitem[\protect\citeauthoryear{{Asgari} et~al.,}{{Asgari}
  et~al.}{2021}]{asgari20}
{Asgari} M.,  et~al., 2021, \mn@doi [\aap] {10.1051/0004-6361/202039070}, \href
  {https://ui.adsabs.harvard.edu/abs/2021A&A...645A.104A} {645, A104}

\bibitem[\protect\citeauthoryear{{Bird}, {Viel}  \& {Haehnelt}}{{Bird}
  et~al.}{2012}]{bird12}
{Bird} S.,  {Viel} M.,   {Haehnelt} M.~G.,  2012, \mn@doi [\mnras]
  {10.1111/j.1365-2966.2011.20222.x}, \href
  {https://ui.adsabs.harvard.edu/abs/2012MNRAS.420.2551B} {420, 2551}

\bibitem[\protect\citeauthoryear{Bishop}{Bishop}{2006}]{bishop06}
Bishop C.~M.,  2006, Pattern Recognition and Machine Learning (Information
  Science and Statistics).
Springer-Verlag, Berlin, Heidelberg

\bibitem[\protect\citeauthoryear{{Blazek}, {Vlah}  \& {Seljak}}{{Blazek}
  et~al.}{2015}]{blazek15}
{Blazek} J.,  {Vlah} Z.,   {Seljak} U.,  2015, \mn@doi [\jcap]
  {10.1088/1475-7516/2015/08/015}, \href
  {http://adsabs.harvard.edu/abs/2015JCAP...08..015B} {8, 015}

\bibitem[\protect\citeauthoryear{{Blazek}, {MacCrann}, {Troxel}  \&
  {Fang}}{{Blazek} et~al.}{2019}]{blazek17}
{Blazek} J.~A.,  {MacCrann} N.,  {Troxel} M.~A.,   {Fang} X.,  2019, \mn@doi
  [\prd] {10.1103/PhysRevD.100.103506}, \href
  {https://ui.adsabs.harvard.edu/abs/2019PhRvD.100j3506B} {100, 103506}

\bibitem[\protect\citeauthoryear{{Bridle} \& {King}}{{Bridle} \&
  {King}}{2007}]{bridle07}
{Bridle} S.,  {King} L.,  2007, \mn@doi [New Journal of Physics]
  {10.1088/1367-2630/9/12/444}, \href
  {http://adsabs.harvard.edu/abs/2007NJPh....9..444B} {9, 444}

\bibitem[\protect\citeauthoryear{{Bridle} et~al.,}{{Bridle}
  et~al.}{2010}]{bridle10}
{Bridle} S.,  et~al., 2010, \mn@doi [\mnras]
  {10.1111/j.1365-2966.2010.16598.x}, \href
  {https://ui.adsabs.harvard.edu/abs/2010MNRAS.405.2044B} {405, 2044}

\bibitem[\protect\citeauthoryear{{Catelan}, {Kamionkowski}  \&
  {Blandford}}{{Catelan} et~al.}{2001}]{catelan01}
{Catelan} P.,  {Kamionkowski} M.,   {Blandford} R.~D.,  2001, \mn@doi [\mnras]
  {10.1046/j.1365-8711.2001.04105.x}, \href
  {http://adsabs.harvard.edu/abs/2001MNRAS.320L...7C} {320, L7}

\bibitem[\protect\citeauthoryear{{Chen} et~al.,}{{Chen} et~al.}{2022}]{chen22}
{Chen} A.,  et~al., 2022, arXiv e-prints, \href
  {https://ui.adsabs.harvard.edu/abs/2022arXiv220608591C} {p. arXiv:2206.08591}

\bibitem[\protect\citeauthoryear{{Crittenden}, {Natarajan}, {Pen}  \&
  {Theuns}}{{Crittenden} et~al.}{2002}]{CrittendenEB}
{Crittenden} R.~G.,  {Natarajan} P.,  {Pen} U.-L.,   {Theuns} T.,  2002,
  \mn@doi [\apj] {10.1086/338838}, \href
  {https://ui.adsabs.harvard.edu/abs/2002ApJ...568...20C} {568, 20}

\bibitem[\protect\citeauthoryear{{DES Collaboration}}{{DES
  Collaboration}}{2022}]{y3-kp}
{DES Collaboration} 2022, \mn@doi [\prd] {10.1103/PhysRevD.105.023520}, \href
  {https://ui.adsabs.harvard.edu/abs/2022PhRvD.105b3520A} {105, 023520}

\bibitem[\protect\citeauthoryear{{Dark Energy Survey Collaboration}}{{Dark
  Energy Survey Collaboration}}{2016}]{sv-cosmicshear}
{Dark Energy Survey Collaboration} 2016, \mn@doi [\prd]
  {10.1103/PhysRevD.94.022001}, \href
  {http://adsabs.harvard.edu/abs/2016PhRvD..94b2001A} {94, 022001}

\bibitem[\protect\citeauthoryear{{DeRose} et~al.,}{{DeRose}
  et~al.}{2019}]{derose19}
{DeRose} J.,  et~al., 2019, \mn@doi [\apj] {10.3847/1538-4357/ab1085}, \href
  {https://ui.adsabs.harvard.edu/abs/2019ApJ...875...69D} {875, 69}

\bibitem[\protect\citeauthoryear{{Desjacques}, {Jeong}  \&
  {Schmidt}}{{Desjacques} et~al.}{2018}]{desjacques18}
{Desjacques} V.,  {Jeong} D.,   {Schmidt} F.,  2018, \mn@doi [\physrep]
  {10.1016/j.physrep.2017.12.002}, \href
  {https://ui.adsabs.harvard.edu/abs/2018PhR...733....1D} {733, 1}

\bibitem[\protect\citeauthoryear{{Doux}, {Baxter}  et~al.}{{Doux}
  et~al.}{2021}]{y3-ppd}
{Doux} C.,  {Baxter} E.,   et~al., 2021, \mn@doi [\mnras]
  {10.1093/mnras/stab526}, \href
  {https://ui.adsabs.harvard.edu/abs/2021MNRAS.503.2688D} {503, 2688}

\bibitem[\protect\citeauthoryear{{Doux} et~al.,}{{Doux}
  et~al.}{2022}]{y3-cosmicshearcls}
{Doux} C.,  et~al., 2022, \mn@doi [\mnras] {10.1093/mnras/stac1826}, \href
  {https://ui.adsabs.harvard.edu/abs/2022MNRAS.tmp.1761D} {}

\bibitem[\protect\citeauthoryear{{Fang}, {Blazek}, {McEwen}  \&
  {Hirata}}{{Fang} et~al.}{2017}]{fang17}
{Fang} X.,  {Blazek} J.~A.,  {McEwen} J.~E.,   {Hirata} C.~M.,  2017, \mn@doi
  [\jcap] {10.1088/1475-7516/2017/02/030}, \href
  {https://ui.adsabs.harvard.edu/abs/2017JCAP...02..030F} {2017, 030}

\bibitem[\protect\citeauthoryear{{Feroz}, {Hobson}, {Cameron}  \&
  {Pettitt}}{{Feroz} et~al.}{2019}]{feroz13}
{Feroz} F.,  {Hobson} M.~P.,  {Cameron} E.,   {Pettitt} A.~N.,  2019, \mn@doi
  [The Open Journal of Astrophysics] {10.21105/astro.1306.2144}, \href
  {https://ui.adsabs.harvard.edu/abs/2019OJAp....2E..10F} {2, 10}

\bibitem[\protect\citeauthoryear{{Friedrich} et~al.,}{{Friedrich}
  et~al.}{2021}]{y3-covariance}
{Friedrich} O.,  et~al., 2021, \mn@doi [\mnras] {10.1093/mnras/stab2384}, \href
  {https://ui.adsabs.harvard.edu/abs/2021MNRAS.508.3125F} {508, 3125}

\bibitem[\protect\citeauthoryear{{Hamana} et~al.,}{{Hamana}
  et~al.}{2020}]{hamana20}
{Hamana} T.,  et~al., 2020, \mn@doi [\pasj] {10.1093/pasj/psz138}, \href
  {https://ui.adsabs.harvard.edu/abs/2020PASJ...72...16H} {72, 16}

\bibitem[\protect\citeauthoryear{{Handley}, {Hobson}  \& {Lasenby}}{{Handley}
  et~al.}{2015}]{handley15}
{Handley} W.~J.,  {Hobson} M.~P.,   {Lasenby} A.~N.,  2015, \mn@doi [\mnras]
  {10.1093/mnras/stv1911}, \href
  {https://ui.adsabs.harvard.edu/abs/2015MNRAS.453.4384H} {453, 4384}

\bibitem[\protect\citeauthoryear{{Heymans} et~al.,}{{Heymans}
  et~al.}{2006}]{heymans06}
{Heymans} C.,  et~al., 2006, \mn@doi [\mnras]
  {10.1111/j.1365-2966.2006.10198.x}, \href
  {https://ui.adsabs.harvard.edu/abs/2006MNRAS.368.1323H} {368, 1323}

\bibitem[\protect\citeauthoryear{{Heymans} et~al.,}{{Heymans}
  et~al.}{2013}]{heymans13}
{Heymans} C.,  et~al., 2013, \mn@doi [\mnras] {10.1093/mnras/stt601}, \href
  {https://ui.adsabs.harvard.edu/abs/2013MNRAS.432.2433H} {432, 2433}

\bibitem[\protect\citeauthoryear{{Hikage} et~al.,}{{Hikage}
  et~al.}{2019}]{hikage19}
{Hikage} C.,  et~al., 2019, \mn@doi [\pasj] {10.1093/pasj/psz010}, \href
  {https://ui.adsabs.harvard.edu/abs/2019PASJ...71...43H} {71, 43}

\bibitem[\protect\citeauthoryear{{Hildebrandt} et~al.,}{{Hildebrandt}
  et~al.}{2017}]{hildebrandt17}
{Hildebrandt} H.,  et~al., 2017, \mn@doi [\mnras] {10.1093/mnras/stw2805},
  \href {https://ui.adsabs.harvard.edu/abs/2017MNRAS.465.1454H} {465, 1454}

\bibitem[\protect\citeauthoryear{{Hirata} \& {Seljak}}{{Hirata} \&
  {Seljak}}{2004}]{hirata04}
{Hirata} C.~M.,  {Seljak} U.,  2004, \mn@doi [\prd]
  {10.1103/PhysRevD.70.063526}, \href
  {http://adsabs.harvard.edu/abs/2004PhRvD..70f3526H} {70, 063526}

\bibitem[\protect\citeauthoryear{Hirata \& Seljak}{Hirata \&
  Seljak}{2010}]{hirata10}
Hirata C.~M.,  Seljak U. c.~v.,  2010, \mn@doi [Phys. Rev. D]
  {10.1103/PhysRevD.82.049901}, 82, 049901

\bibitem[\protect\citeauthoryear{{Hirata}, {Mandelbaum}, {Ishak}, {Seljak},
  {Nichol}, {Pimbblet}, {Ross}  \& {Wake}}{{Hirata} et~al.}{2007}]{hirata07}
{Hirata} C.~M.,  {Mandelbaum} R.,  {Ishak} M.,  {Seljak} U.,  {Nichol} R.,
  {Pimbblet} K.~A.,  {Ross} N.~P.,   {Wake} D.,  2007, \mn@doi [\mnras]
  {10.1111/j.1365-2966.2007.12312.x}, \href
  {https://ui.adsabs.harvard.edu/abs/2007MNRAS.381.1197H} {381, 1197}

\bibitem[\protect\citeauthoryear{{Ivezi{\'c}} et~al.,}{{Ivezi{\'c}}
  et~al.}{2019}]{lsst2019}
{Ivezi{\'c}} {\v{Z}}.,  et~al., 2019, \mn@doi [\apj]
  {10.3847/1538-4357/ab042c}, \href
  {https://ui.adsabs.harvard.edu/abs/2019ApJ...873..111I} {873, 111}

\bibitem[\protect\citeauthoryear{{Jarvis} et~al.,}{{Jarvis}
  et~al.}{2021}]{jarvis21}
{Jarvis} M.,  et~al., 2021, \mn@doi [\mnras] {10.1093/mnras/staa3679}, \href
  {https://ui.adsabs.harvard.edu/abs/2021MNRAS.501.1282J} {501, 1282}

\bibitem[\protect\citeauthoryear{{Jee}, {Tyson}, {Hilbert}, {Schneider},
  {Schmidt}  \& {Wittman}}{{Jee} et~al.}{2016}]{DLS}
{Jee} M.~J.,  {Tyson} J.~A.,  {Hilbert} S.,  {Schneider} M.~D.,  {Schmidt} S.,
   {Wittman} D.,  2016, \mn@doi [\apj] {10.3847/0004-637X/824/2/77}, \href
  {https://ui.adsabs.harvard.edu/abs/2016ApJ...824...77J} {824, 77}

\bibitem[\protect\citeauthoryear{Jeffreys}{Jeffreys}{1961}]{jeffreys68}
Jeffreys H.,  1961, The theory of probability.
OUP Oxford

\bibitem[\protect\citeauthoryear{{Joachimi} et~al.,}{{Joachimi}
  et~al.}{2015}]{joachimi15}
{Joachimi} B.,  et~al., 2015, \mn@doi [\ssr] {10.1007/s11214-015-0177-4}, \href
  {https://ui.adsabs.harvard.edu/abs/2015SSRv..193....1J} {193, 1}

\bibitem[\protect\citeauthoryear{{Joachimi} et~al.,}{{Joachimi}
  et~al.}{2021a}]{joachimi21}
{Joachimi} B.,  et~al., 2021a, \mn@doi [\aap] {10.1051/0004-6361/202038831},
  \href {https://ui.adsabs.harvard.edu/abs/2021A&A...646A.129J} {646, A129}

\bibitem[\protect\citeauthoryear{{Joachimi}, {K{\"o}hlinger}, {Handley}  \&
  {Lemos}}{{Joachimi} et~al.}{2021b}]{joachimi21a}
{Joachimi} B.,  {K{\"o}hlinger} F.,  {Handley} W.,   {Lemos} P.,  2021b,
  \mn@doi [\aap] {10.1051/0004-6361/202039560}, \href
  {https://ui.adsabs.harvard.edu/abs/2021A&A...647L...5J} {647, L5}

\bibitem[\protect\citeauthoryear{{Johnston} et~al.,}{{Johnston}
  et~al.}{2019}]{johnston19}
{Johnston} H.,  et~al., 2019, \mn@doi [\aap] {10.1051/0004-6361/201834714},
  \href {https://ui.adsabs.harvard.edu/abs/2019A&A...624A..30J} {624, A30}

\bibitem[\protect\citeauthoryear{{Kass} \& {Raftery}}{{Kass} \&
  {Raftery}}{1995}]{kass95}
{Kass} R.,  {Raftery} A.,  1995, \mn@doi [Journal of the American Statistical
  Association] {10.1080/01621459.1995.10476572}, 90, 773

\bibitem[\protect\citeauthoryear{Kerscher \& Weller}{Kerscher \&
  Weller}{2019}]{Kerscher19}
Kerscher M.,  Weller J.,  2019, \mn@doi [SciPost Phys. Lect. Notes]
  {10.21468/SciPostPhysLectNotes.9}, p.~9

\bibitem[\protect\citeauthoryear{{Kiessling} et~al.,}{{Kiessling}
  et~al.}{2015}]{kiessling15}
{Kiessling} A.,  et~al., 2015, \mn@doi [\ssr] {10.1007/s11214-015-0203-6},
  \href {https://ui.adsabs.harvard.edu/abs/2015SSRv..193...67K} {193, 67}

\bibitem[\protect\citeauthoryear{{Kilbinger} et~al.,}{{Kilbinger}
  et~al.}{2010}]{kilbinger10}
{Kilbinger} M.,  et~al., 2010, \mn@doi [\mnras]
  {10.1111/j.1365-2966.2010.16605.x}, \href
  {https://ui.adsabs.harvard.edu/abs/2010MNRAS.405.2381K} {405, 2381}

\bibitem[\protect\citeauthoryear{{Kirk} et~al.,}{{Kirk} et~al.}{2015}]{kirk15}
{Kirk} D.,  et~al., 2015, \mn@doi [\ssr] {10.1007/s11214-015-0213-4}, \href
  {https://ui.adsabs.harvard.edu/abs/2015SSRv..193..139K} {193, 139}

\bibitem[\protect\citeauthoryear{{Knabenhans} et~al.,}{{Knabenhans}
  et~al.}{2021}]{knabenhans21}
{Knabenhans} M.,  et~al., 2021, \mn@doi [\mnras] {10.1093/mnras/stab1366},
  \href {https://ui.adsabs.harvard.edu/abs/2021MNRAS.505.2840E} {505, 2840}

\bibitem[\protect\citeauthoryear{{Krause} et~al.,}{{Krause}
  et~al.}{2017}]{krause17}
{Krause} E.,  et~al., 2017, arXiv e-prints, \href
  {https://ui.adsabs.harvard.edu/abs/2017arXiv170609359K} {p. arXiv:1706.09359}

\bibitem[\protect\citeauthoryear{{Krause} et~al.,}{{Krause}
  et~al.}{2021}]{krause21}
{Krause} E.,  et~al., 2021, arXiv e-prints, \href
  {https://ui.adsabs.harvard.edu/abs/2021arXiv210513548K} {p. arXiv:2105.13548}

\bibitem[\protect\citeauthoryear{{Laureijs} et~al.,}{{Laureijs}
  et~al.}{2011}]{euclid}
{Laureijs} R.,  et~al., 2011, arXiv e-prints, \href
  {https://ui.adsabs.harvard.edu/abs/2011arXiv1110.3193L} {p. arXiv:1110.3193}

\bibitem[\protect\citeauthoryear{{Lemos}, {Raveri}, {Campos}  et~al.}{{Lemos}
  et~al.}{2021}]{y3-tensions}
{Lemos} P.,  {Raveri} M.,  {Campos} A.,   et~al., 2021, \mn@doi [\mnras]
  {10.1093/mnras/stab1670}, \href
  {https://ui.adsabs.harvard.edu/abs/2021MNRAS.505.6179L} {505, 6179}

\bibitem[\protect\citeauthoryear{{Lemos}, {Weaverdyck}  et~al.}{{Lemos}
  et~al.}{2022}]{y3-samplers}
{Lemos} P.,  {Weaverdyck} N.,   et~al., 2022, \mn@doi [\mnras]
  {10.1093/mnras/stac2786}, \href
  {https://ui.adsabs.harvard.edu/abs/2022MNRAS.tmp.2714L} {}

\bibitem[\protect\citeauthoryear{{Lewis}}{{Lewis}}{2019}]{lewis19}
{Lewis} A.,  2019, arXiv e-prints, \href
  {https://ui.adsabs.harvard.edu/abs/2019arXiv191013970L} {p. arXiv:1910.13970}

\bibitem[\protect\citeauthoryear{{Lewis} \& {Bridle}}{{Lewis} \&
  {Bridle}}{2002}]{lewis02}
{Lewis} A.,  {Bridle} S.,  2002, \mn@doi [\prd] {10.1103/PhysRevD.66.103511},
  \href {https://ui.adsabs.harvard.edu/abs/2002PhRvD..66j3511L} {66, 103511}

\bibitem[\protect\citeauthoryear{{Lewis}, {Challinor}  \& {Lasenby}}{{Lewis}
  et~al.}{2000}]{lewis00}
{Lewis} A.,  {Challinor} A.,   {Lasenby} A.,  2000, \mn@doi [\apj]
  {10.1086/309179}, \href
  {https://ui.adsabs.harvard.edu/abs/2000ApJ...538..473L} {538, 473}

\bibitem[\protect\citeauthoryear{{Liddle}}{{Liddle}}{2007}]{liddle07}
{Liddle} A.~R.,  2007, \mn@doi [\mnras] {10.1111/j.1745-3933.2007.00306.x},
  \href {https://ui.adsabs.harvard.edu/abs/2007MNRAS.377L..74L} {377, L74}

\bibitem[\protect\citeauthoryear{Liddle, Mukherjee  \& Parkinson}{Liddle
  et~al.}{2006a}]{liddle06}
Liddle A.,  Mukherjee P.,   Parkinson D.,  2006a, \mn@doi [Astronomy &
  Geophysics] {10.1111/j.1468-4004.2006.47430.x}, 47, 4.30

\bibitem[\protect\citeauthoryear{{Liddle}, {Mukherjee}, {Parkinson}  \&
  {Wang}}{{Liddle} et~al.}{2006b}]{liddle06b}
{Liddle} A.~R.,  {Mukherjee} P.,  {Parkinson} D.,   {Wang} Y.,  2006b, \mn@doi
  [\prd] {10.1103/PhysRevD.74.123506}, \href
  {https://ui.adsabs.harvard.edu/abs/2006PhRvD..74l3506L} {74, 123506}

\bibitem[\protect\citeauthoryear{{Limber}}{{Limber}}{1953}]{Limber53}
{Limber} D.~N.,  1953, \mn@doi [\apj] {10.1086/145672}, \href
  {https://ui.adsabs.harvard.edu/abs/1953ApJ...117..134L} {117, 134}

\bibitem[\protect\citeauthoryear{LoVerde \& Afshordi}{LoVerde \&
  Afshordi}{2008}]{Limber_LoVerde2008}
LoVerde M.,  Afshordi N.,  2008, \mn@doi [Phys. Rev. D]
  {10.1103/PhysRevD.78.123506}, 78, 123506

\bibitem[\protect\citeauthoryear{{Loureiro} et~al.,}{{Loureiro}
  et~al.}{2022}]{loureiro21}
{Loureiro} A.,  et~al., 2022, \mn@doi [\aap] {10.1051/0004-6361/202142481},
  \href {https://ui.adsabs.harvard.edu/abs/2022A&A...665A..56L} {665, A56}

\bibitem[\protect\citeauthoryear{{Mandelbaum} et~al.,}{{Mandelbaum}
  et~al.}{2015}]{mandelbaum15}
{Mandelbaum} R.,  et~al., 2015, \mn@doi [\mnras] {10.1093/mnras/stv781}, \href
  {https://ui.adsabs.harvard.edu/abs/2015MNRAS.450.2963M} {450, 2963}

\bibitem[\protect\citeauthoryear{{Marshall}, {Rajguru}  \& {Slosar}}{{Marshall}
  et~al.}{2006}]{marshall06}
{Marshall} P.,  {Rajguru} N.,   {Slosar} A.,  2006, \mn@doi [\prd]
  {10.1103/PhysRevD.73.067302}, \href
  {https://ui.adsabs.harvard.edu/abs/2006PhRvD..73f7302M} {73, 067302}

\bibitem[\protect\citeauthoryear{{McEwen}, {Fang}, {Hirata}  \&
  {Blazek}}{{McEwen} et~al.}{2016}]{mcewen16}
{McEwen} J.~E.,  {Fang} X.,  {Hirata} C.~M.,   {Blazek} J.~A.,  2016, \mn@doi
  [\jcap] {10.1088/1475-7516/2016/09/015}, \href
  {https://ui.adsabs.harvard.edu/abs/2016JCAP...09..015M} {2016, 015}

\bibitem[\protect\citeauthoryear{{Mead}, {Brieden}, {Tr{\"o}ster}  \&
  {Heymans}}{{Mead} et~al.}{2021}]{mead21}
{Mead} A.~J.,  {Brieden} S.,  {Tr{\"o}ster} T.,   {Heymans} C.,  2021, \mn@doi
  [\mnras] {10.1093/mnras/stab082}, \href
  {https://ui.adsabs.harvard.edu/abs/2021MNRAS.502.1401M} {502, 1401}

\bibitem[\protect\citeauthoryear{{Myles}, {Alarcon}  et~al.}{{Myles}
  et~al.}{2021}]{y3-sompz}
{Myles} J.,  {Alarcon} A.,   et~al., 2021, \mn@doi [\mnras]
  {10.1093/mnras/stab1515}, \href
  {https://ui.adsabs.harvard.edu/abs/2021MNRAS.505.4249M} {505, 4249}

\bibitem[\protect\citeauthoryear{{Neal}}{{Neal}}{1998}]{neal98}
{Neal} R.~M.,  1998, arXiv e-prints, \href
  {https://ui.adsabs.harvard.edu/abs/1998physics...3008N} {p. physics/9803008}

\bibitem[\protect\citeauthoryear{{Osato}, {Shirasaki}  \& {Yoshida}}{{Osato}
  et~al.}{2015}]{osato15}
{Osato} K.,  {Shirasaki} M.,   {Yoshida} N.,  2015, \mn@doi [\apj]
  {10.1088/0004-637X/806/2/186}, \href
  {https://ui.adsabs.harvard.edu/abs/2015ApJ...806..186O} {806, 186}

\bibitem[\protect\citeauthoryear{{Padilla}, {Tellez}, {Escamilla}  \&
  {Vazquez}}{{Padilla} et~al.}{2019}]{padilla19}
{Padilla} L.~E.,  {Tellez} L.~O.,  {Escamilla} L.~A.,   {Vazquez} J.~A.,  2019,
  arXiv e-prints, \href {https://ui.adsabs.harvard.edu/abs/2019arXiv190311127P}
  {p. arXiv:1903.11127}

\bibitem[\protect\citeauthoryear{{Pandey} et~al.,}{{Pandey}
  et~al.}{2020}]{pandey20}
{Pandey} S.,  et~al., 2020, \mn@doi [\prd] {10.1103/PhysRevD.102.123522}, \href
  {https://ui.adsabs.harvard.edu/abs/2020PhRvD.102l3522P} {102, 123522}

\bibitem[\protect\citeauthoryear{{Raveri} \& {Hu}}{{Raveri} \&
  {Hu}}{2019}]{raveri19}
{Raveri} M.,  {Hu} W.,  2019, \mn@doi [\prd] {10.1103/PhysRevD.99.043506},
  \href {https://ui.adsabs.harvard.edu/abs/2019PhRvD..99d3506R} {99, 043506}

\bibitem[\protect\citeauthoryear{Rigdon}{Rigdon}{1999}]{rigdon99}
Rigdon E.~E.,  1999, \mn@doi [Structural Equation Modeling: A Multidisciplinary
  Journal] {10.1080/10705519909540131}, 6, 219

\bibitem[\protect\citeauthoryear{{Saito}, {Takada}  \& {Taruya}}{{Saito}
  et~al.}{2008}]{saito08}
{Saito} S.,  {Takada} M.,   {Taruya} A.,  2008, \mn@doi [\prl]
  {10.1103/PhysRevLett.100.191301}, \href
  {https://ui.adsabs.harvard.edu/abs/2008PhRvL.100s1301S} {100, 191301}

\bibitem[\protect\citeauthoryear{{Samuroff} et~al.,}{{Samuroff}
  et~al.}{2019}]{y1-tatt}
{Samuroff} S.,  et~al., 2019, \mn@doi [\mnras] {10.1093/mnras/stz2197}, \href
  {https://ui.adsabs.harvard.edu/abs/2019MNRAS.489.5453S} {489, 5453}

\bibitem[\protect\citeauthoryear{{S{\'a}nchez}, {Prat}  et~al.}{{S{\'a}nchez}
  et~al.}{2022}]{y3-shearratio}
{S{\'a}nchez} C.,  {Prat} J.,   et~al., 2022, \mn@doi [\prd]
  {10.1103/PhysRevD.105.083529}, \href
  {https://ui.adsabs.harvard.edu/abs/2022PhRvD.105h3529S} {105, 083529}

\bibitem[\protect\citeauthoryear{Schermelleh-Engel, Moosbrugger  \&
  Müller}{Schermelleh-Engel et~al.}{2003}]{engel03}
Schermelleh-Engel K.,  Moosbrugger H.,   Müller H.,  2003, Methods of
  Psychological Research Online, 8, 23–74

\bibitem[\protect\citeauthoryear{{Schneider}, {van Waerbeke}  \&
  {Mellier}}{{Schneider} et~al.}{2002}]{Schneider_vanW_Mell_2002}
{Schneider} P.,  {van Waerbeke} L.,   {Mellier} Y.,  2002, \mn@doi [\aap]
  {10.1051/0004-6361:20020626}, \href
  {https://ui.adsabs.harvard.edu/abs/2002A&A...389..729S} {389, 729}

\bibitem[\protect\citeauthoryear{{Schwarz}}{{Schwarz}}{1978}]{schwarz78}
{Schwarz} G.,  1978, Annals of Statistics, 6, 461–464

\bibitem[\protect\citeauthoryear{{Secco}, {Samuroff}  et~al.}{{Secco}
  et~al.}{2022}]{y3-cosmicshear2}
{Secco} L.~F.,  {Samuroff} S.,   et~al., 2022, \mn@doi [\prd]
  {10.1103/PhysRevD.105.023515}, \href
  {https://ui.adsabs.harvard.edu/abs/2022PhRvD.105b3515S} {105, 023515}

\bibitem[\protect\citeauthoryear{{Simon} \& {Hilbert}}{{Simon} \&
  {Hilbert}}{2018}]{simon18}
{Simon} P.,  {Hilbert} S.,  2018, \mn@doi [\aap] {10.1051/0004-6361/201732248},
  \href {https://ui.adsabs.harvard.edu/abs/2018A&A...613A..15S} {613, A15}

\bibitem[\protect\citeauthoryear{{Singh} \& {Mandelbaum}}{{Singh} \&
  {Mandelbaum}}{2016}]{singh16}
{Singh} S.,  {Mandelbaum} R.,  2016, \mn@doi [\mnras] {10.1093/mnras/stw144},
  \href {https://ui.adsabs.harvard.edu/abs/2016MNRAS.457.2301S} {457, 2301}

\bibitem[\protect\citeauthoryear{Skilling}{Skilling}{2006}]{Skilling:2006}
Skilling J.,  2006, \mn@doi [Bayesian Anal.] {10.1214/06-BA127}, 1, 833

\bibitem[\protect\citeauthoryear{{Spergel} et~al.,}{{Spergel}
  et~al.}{2015}]{roman}
{Spergel} D.,  et~al., 2015, arXiv e-prints, \href
  {https://ui.adsabs.harvard.edu/abs/2015arXiv150303757S} {p. arXiv:1503.03757}

\bibitem[\protect\citeauthoryear{{Steiger}, {Shapiro}  \& {Browne}}{{Steiger}
  et~al.}{1985}]{steiger85}
{Steiger} J.,  {Shapiro} A.,   {Browne} M.,  1985, \mn@doi [Psychometrika]
  {10.1007/BF02294104}, 50, 253

\bibitem[\protect\citeauthoryear{{Takahashi}, {Sato}, {Nishimichi}, {Taruya}
  \& {Oguri}}{{Takahashi} et~al.}{2012}]{takahashi12}
{Takahashi} R.,  {Sato} M.,  {Nishimichi} T.,  {Taruya} A.,   {Oguri} M.,
  2012, \mn@doi [\apj] {10.1088/0004-637X/761/2/152}, \href
  {https://ui.adsabs.harvard.edu/abs/2012ApJ...761..152T} {761, 152}

\bibitem[\protect\citeauthoryear{Tokdar \& Kass}{Tokdar \&
  Kass}{2010}]{todkar10}
Tokdar S.~T.,  Kass R.~E.,  2010, \mn@doi [WIREs Computational Statistics]
  {https://doi.org/10.1002/wics.56}, 2, 54

\bibitem[\protect\citeauthoryear{{Tr{\"o}ster} et~al.,}{{Tr{\"o}ster}
  et~al.}{2022}]{troster22}
{Tr{\"o}ster} T.,  et~al., 2022, \mn@doi [\aap] {10.1051/0004-6361/202142197},
  \href {https://ui.adsabs.harvard.edu/abs/2022A&A...660A..27T} {660, A27}

\bibitem[\protect\citeauthoryear{{Trotta}}{{Trotta}}{2007}]{trotta07}
{Trotta} R.,  2007, \mn@doi [\mnras] {10.1111/j.1365-2966.2007.11738.x}, \href
  {https://ui.adsabs.harvard.edu/abs/2007MNRAS.378...72T} {378, 72}

\bibitem[\protect\citeauthoryear{Trotta}{Trotta}{2008}]{trotta08}
Trotta R.,  2008, \mn@doi [Contemporary Physics] {10.1080/00107510802066753},
  49, 71

\bibitem[\protect\citeauthoryear{{Troxel} \& {Ishak}}{{Troxel} \&
  {Ishak}}{2015}]{troxel15}
{Troxel} M.~A.,  {Ishak} M.,  2015, \mn@doi [\physrep]
  {10.1016/j.physrep.2014.11.001}, \href
  {https://ui.adsabs.harvard.edu/abs/2015PhR...558....1T} {558, 1}

\bibitem[\protect\citeauthoryear{{Troxel} et~al.,}{{Troxel}
  et~al.}{2018}]{y1-cosmicshear}
{Troxel} M.~A.,  et~al., 2018, \mn@doi [\prd] {10.1103/PhysRevD.98.043528},
  \href {https://ui.adsabs.harvard.edu/abs/2018PhRvD..98d3528T} {98, 043528}

\bibitem[\protect\citeauthoryear{{Vardanyan}, {Trotta}  \& {Silk}}{{Vardanyan}
  et~al.}{2011}]{vardanyan11}
{Vardanyan} M.,  {Trotta} R.,   {Silk} J.,  2011, \mn@doi [\mnras]
  {10.1111/j.1745-3933.2011.01040.x}, \href
  {https://ui.adsabs.harvard.edu/abs/2011MNRAS.413L..91V} {413, L91}

\bibitem[\protect\citeauthoryear{{Weaverdyck}, {Alves}  et~al.}{{Weaverdyck}
  et~al.}{2022}]{weaverdyck22}
{Weaverdyck} N.,  {Alves} O.,   et~al., 2022, in prep

\bibitem[\protect\citeauthoryear{Wilks}{Wilks}{1938}]{wilks38}
Wilks S.~S.,  1938, \mn@doi [The Annals of Mathematical Statistics]
  {10.1214/aoms/1177732360}, 9, 60

\bibitem[\protect\citeauthoryear{{Zuntz} et~al.,}{{Zuntz}
  et~al.}{2015}]{zuntz15}
{Zuntz} J.,  et~al., 2015, \mn@doi [Astronomy and Computing]
  {10.1016/j.ascom.2015.05.005}, \href
  {https://ui.adsabs.harvard.edu/abs/2015A&C....12...45Z} {12, 45}

\makeatother
\end{thebibliography}

\appendix

\section{Parameters \& Priors}
\label{sec:appendixA}

Our setup matches the fiducial choices of the DES Y3 cosmic shear analysis. The only significant difference is that, for the sake of simplicity, we choose not to use the additional shear ratio likelihood included by \citet*{y3-cosmicshear2};  \citet{y3-cosmicshear1} (a similar decision was made for validating the analysis choices pre-unblinding; see \citealt{krause21}). As a result, our model space is slightly smaller, since we do not need to vary parameters for galaxy bias or lens photo$-z$ error. The corresponding parameters and their priors are shown in Table \ref{table: priors}. Note that these are \emph{almost} identical to the priors used in the Y3 analysis, except for those on the shear calibration parameters, which have been shifted to match the input to the simulated data.

\begin{figure}
\includegraphics[width=\columnwidth]{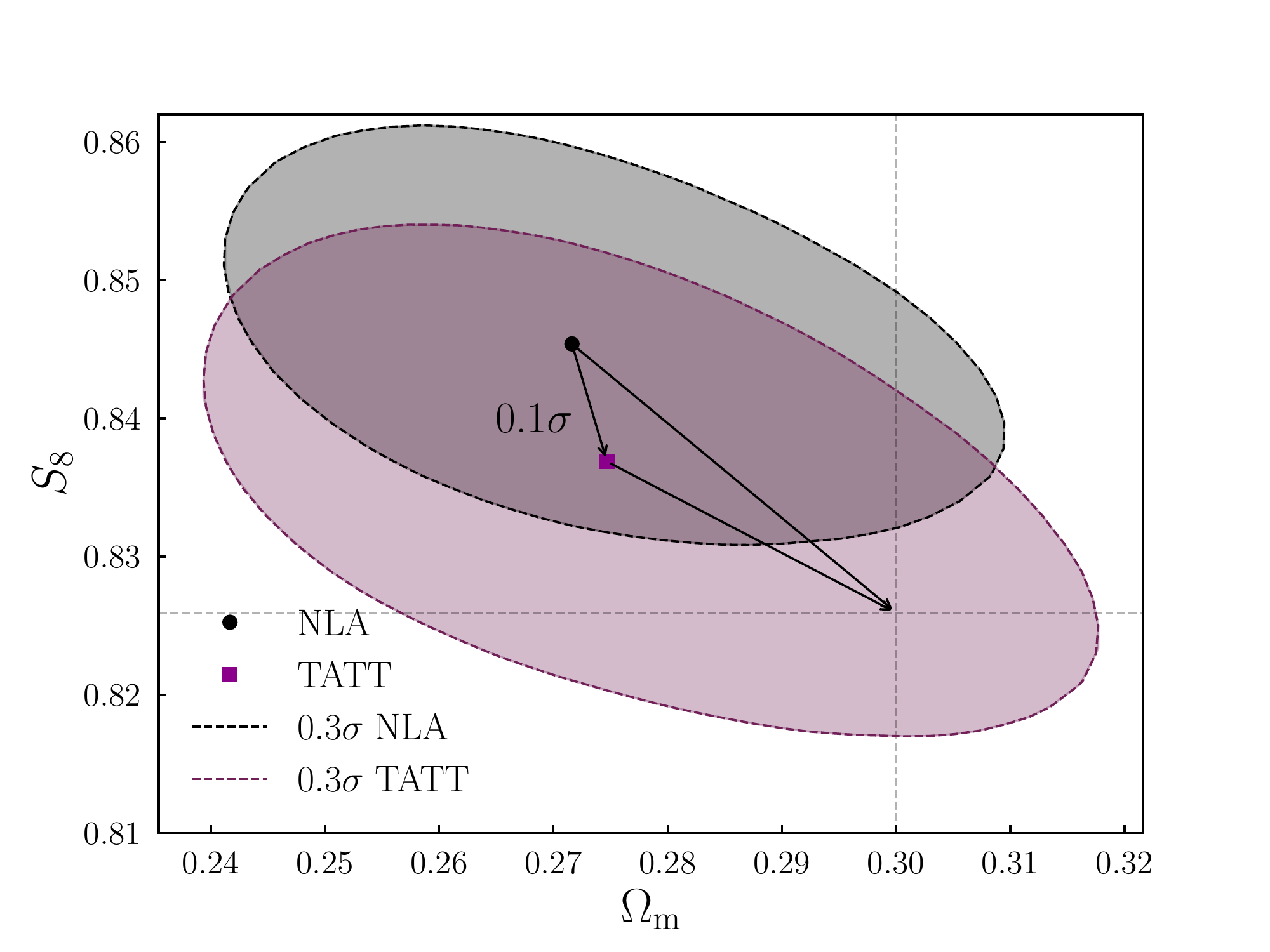}
\caption{Projected $0.3\sigma$ contours from NLA and TATT chains run on a noisy NLA data vector (see Section~\ref{sec:method:bias} for definitions). The NLA input parameters are $A_1=0.7$, $\eta_1=-1.7$. Since, by construction, both IA models are sufficient to describe the data, any residual offset is thought to be the result of projection effects. As labelled, this is at the level of $0.1\sigma$ for our analysis setup. }
\label{fig:bias_nla_dv}
\end{figure}

In a particular setup, one should expect some level of projection effects in the marginal parameter constraints \citep{krause21}. Since such offsets are artefacts of the way we choose to visualise our results (i.e., the global best fit is still accurate) it is not, in general, useful to think of them as a form of bias; our method does, however, rely on our ability to interpret differences in the 2D projected $S_8-\omegam$ plane. It is thus helpful to try to quantify such effects in our case. 
In Figure~\ref{fig:bias_nla_dv} we show the results of our NLA and TATT analyses on an NLA-only data vector, with our fiducial noise realisation. That is, in this case, both the TATT and NLA models can reproduce the data exactly (up to noise).
The offset between the best-fitting parameters when using the two models, shown in Figure~\ref{fig:bias_nla_dv}, is at the level of $\sim 0.1\sigma$. This effectively provides a floor to the bias in our analysis. Although we can occasionally find biases below this level due to noise (see Section~\ref{sec:results} for discussion), we should consider all these cases as unbiased, at least to within the uncertainty due to projection effects.
Note that this is consistent with the results of \citet{krause21}, who performed a similar test using noiseless data (see their Figure~4). 

Note that projection effects are complicated, and may be a function of (among other things) the choice of input parameters, noise realisation, priors and constraining power of the data. Although it is reassuring that our result matches that of \citet{krause21}, there may still be some residual uncertainty in the size of the effect. This is not, however, necessarily a problem for our method. Indeed, variable projection effects would simply add an extra source of noise in the bias-metric relation, which would be factored into our results in the same way as, e.g., chain-to-chain sampler noise.

\begin{table}
	\centering
	\vspace{1cm}
    \caption{A summary of the central values and priors used in our analysis. The top seven rows are cosmological parameters, while those in the lower sections are nuisance parameters corresponding to astrophysics and data calibration. Priors are either uniform (U) or normally-distributed, $\mathcal{N}(\mu,\sigma)$. Note the IA parameters are marked with a star because many different values are used as input to our data vectors, as discussed in Section \ref{sec:data:ia_sampling}. The values shown here are used for convenience, whenever it is useful to show/discuss a single data realisation (e.g., in Figure \ref{fig:noisy_data_vec}). 
    } \label{table: priors}

\begin{tabular}{c c c }
\hline 
\hline
\bf{Parameter} & \bf{Fiducial Value} & \bf{Prior}\tabularnewline
\hline 
\hline 
\multicolumn{3}{c}{{\bf Cosmological  Parameters}} \\
\omegam & $0.29$ & $\mathrm{U}[0.1, 0.9]$ \tabularnewline
\omegab & $0.052$ & $\mathrm{U} [0.03, 0.07]$ \tabularnewline
$h$     & $0.75$ & $\mathrm{U}[0.55, 0.91]$ \tabularnewline
\as     & $2.38 \times 10^{-9}$ & $\mathrm{U}[0.5,  5.0]\times 10^{-9}$ \tabularnewline
\ns     & $0.99$ & $\mathrm{U}[0.87, 1.07]$   \tabularnewline
$\Omega_{\nu}h^2$ & $0.00053$ & $\mathrm{U}[0.6, 6.44]\times 10^{-3}$   \tabularnewline
\hline
\multicolumn{3}{c}{{\bf Calibration  Parameters}}  \\
$m_{1}$ & $0.0$ & $\mathcal{N}(0.0, 0.0059)$ \tabularnewline
$m_{2}$ & $0.0$ & $\mathcal{N}(0.0, 0.0042)$ \tabularnewline
$m_{3}$ & $0.0$ & $\mathcal{N}(0.0,  0.0054)$ \tabularnewline
$m_{4}$ & $0.0$ & $\mathcal{N}(0.0, 0.0072)$ \tabularnewline
$\Delta z_{1}$ & $0.0$ & $\mathcal{N}(0.0,0.018)$ \tabularnewline
$\Delta z_{2}$ & $0.0$ & $\mathcal{N}(0.0,0.015)$ \tabularnewline
$\Delta z_{3}$ & $0.0$ & $\mathcal{N}(0.0,0.011)$ \tabularnewline
$\Delta z_{4}$ & $0.0$ & $\mathcal{N}(0.0, 0.017)$ \tabularnewline
\hline
\multicolumn{3}{c}{{\bf Intrinsic Alignment Parameters *}} \\
$A_1$ & $0.7$ & $\mathrm{U}[-5, 5]$ \tabularnewline
$A_2$ & $-1.36$ & $\mathrm{U}[-5, 5]$ \tabularnewline
$\eta_1$ & $-1.7$ & $\mathrm{U}[-5, 5]$ \tabularnewline
$\eta_2$ & $-2.5$ & $\mathrm{U}[-5, 5]$ \tabularnewline
$b_{\rm TA}$ & $1.0$ & $\mathrm{U}[0, 2]$ \tabularnewline
\hline 
\hline 
\end{tabular}

\end{table}

\section{NLA \& TATT Posteriors}\label{app:posteriors}

\begin{figure}
\includegraphics[width=\columnwidth]{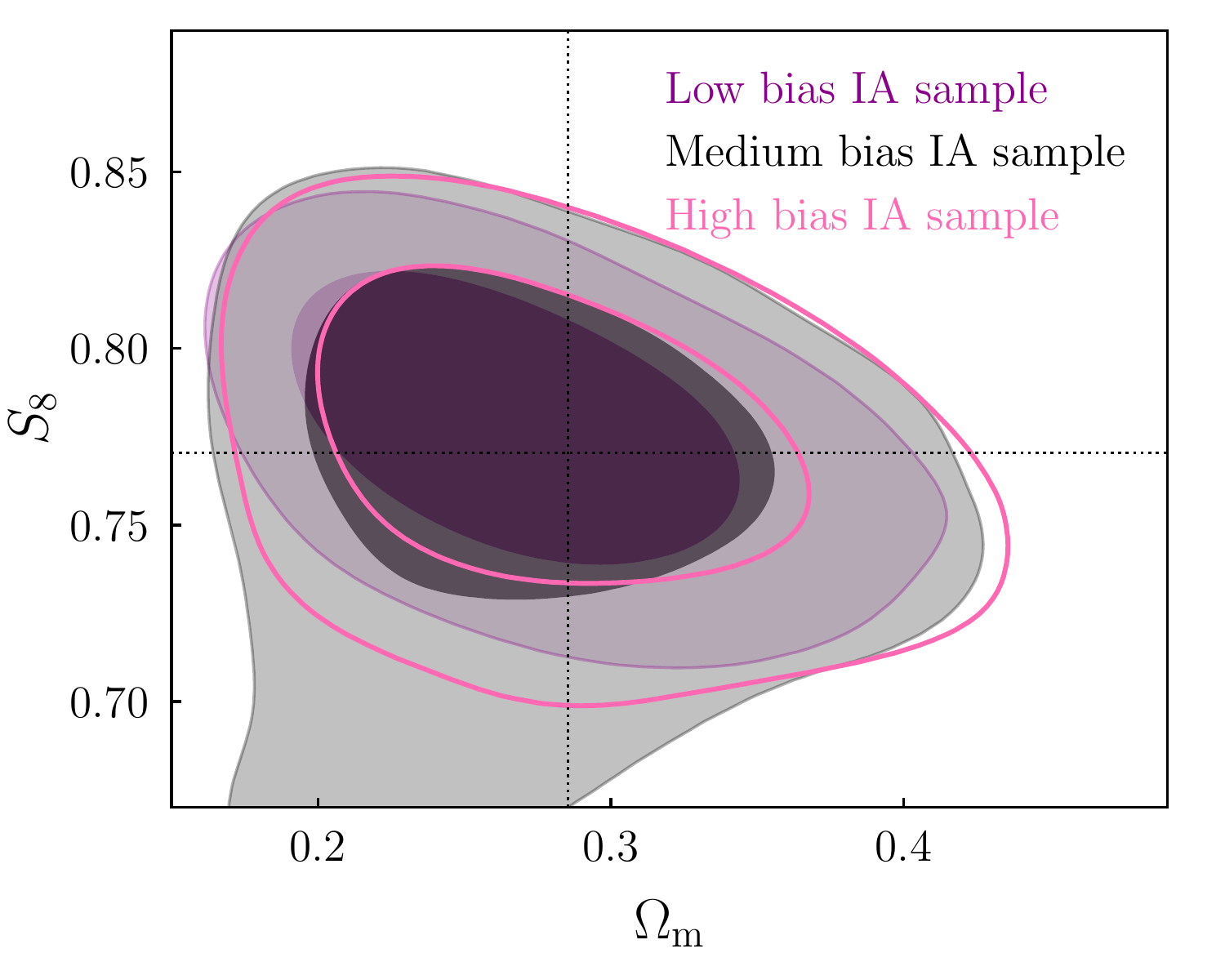}
\includegraphics[width=\columnwidth]{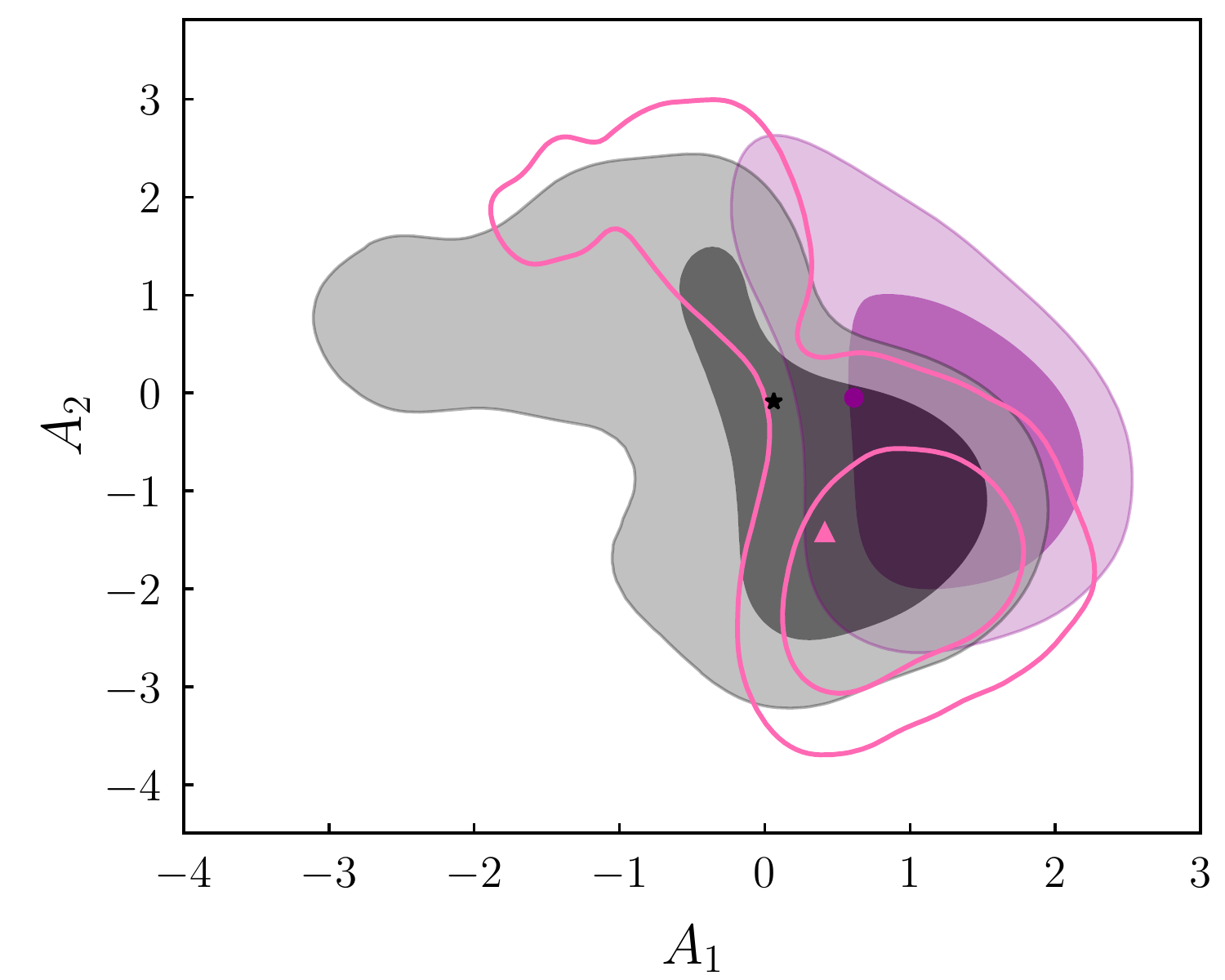}
\caption{\textbf{Top:} 68\% and 95\% cosmology confidence contours from TATT model fits on simulated noisy data vectors. Like in Figure~\ref{fig:nla_contours}, the different colours represent samples selected to cover a range from relatively extreme (i.e., large bias in NLA) to mild (low bias) cases. The dotted cross represents the input cosmological parameters (which is offset from the center of the contours due to data vector noise).
\textbf{Bottom:} The same, but showing the two TATT amplitude parameters. The markers (dot, star, triangle) show the input IA parameters for each case.
}
\label{fig:contours_tatt_nla}
\end{figure}

For completeness, in Figure~\ref{fig:contours_tatt_nla} we show the TATT model posteriors for the IA scenarios discussed in Section~\ref{sec:results} and Figure~\ref{fig:nla_contours}. In that section we discussed three sets of IA model parameters that were selected to give a range of severity of $S_8-\omegam$ bias in NLA. Our results there showed that significant biases can be present in an NLA analysis without necessarily distorting the shape of the contours or giving a ``bad" $\chi^2$ (interpreted in the conventional way, using statistically-motivated cut-offs). 
As expected, the cosmological parameter contours in Figure~\ref{fig:contours_tatt_nla} (upper panel) are consistent with each other. Since the data vectors contain (the same) noise, they are offset from the input point slightly. Depending on the input scenario, the width also varies slightly, primarily due to the tail in $A_1-A_2$ space, which correlates with $S_8$.

It is worth also briefly commenting here on the shapes of the TATT posteriors. It has been observed before that the TATT model can give rise to teardrop shaped, sometimes slightly bimodal contours in the $A_1-A_2$ plane (see for example \citealt*{y3-cosmicshear2} Fig.~8 and \citealt*{y3-shearratio} Fig.~15). A significant tail to positive (or negative) $A_2$ tends to create a tail in $S_8$, of the sort seen in grey contours in the bottom panel of Figure~\ref{fig:contours_tatt_nla}. Note however that the shape and asymmetry depends quite heavily on where the posteriors sit in parameter space (meaning the noise realisation as well as the ``true" TATT parameters), and on the constraining power of the data (more constraining power tends to trim away some of the non Gaussian tails). It is not clear that we can use any sort of qualitative assessment based on the TATT posterior shape as an indicator for bias in simpler models.

\section{Bayes Ratio} \label{sec:appendixBayes}

Although we ultimately choose to use the $\Delta \chi^2_{(\textrm{df})}$ as our model comparison statistic, it is also useful to consider other commonly used alternatives. The Bayes ratio has become a popular tool in weak lensing cosmology in recent years, in part because it in principle contains more information than the likelihood. It is also readily available as the by-product of running a nested sampling algorithm to estimate posteriors.

\begin{figure}
\includegraphics[width=\columnwidth]{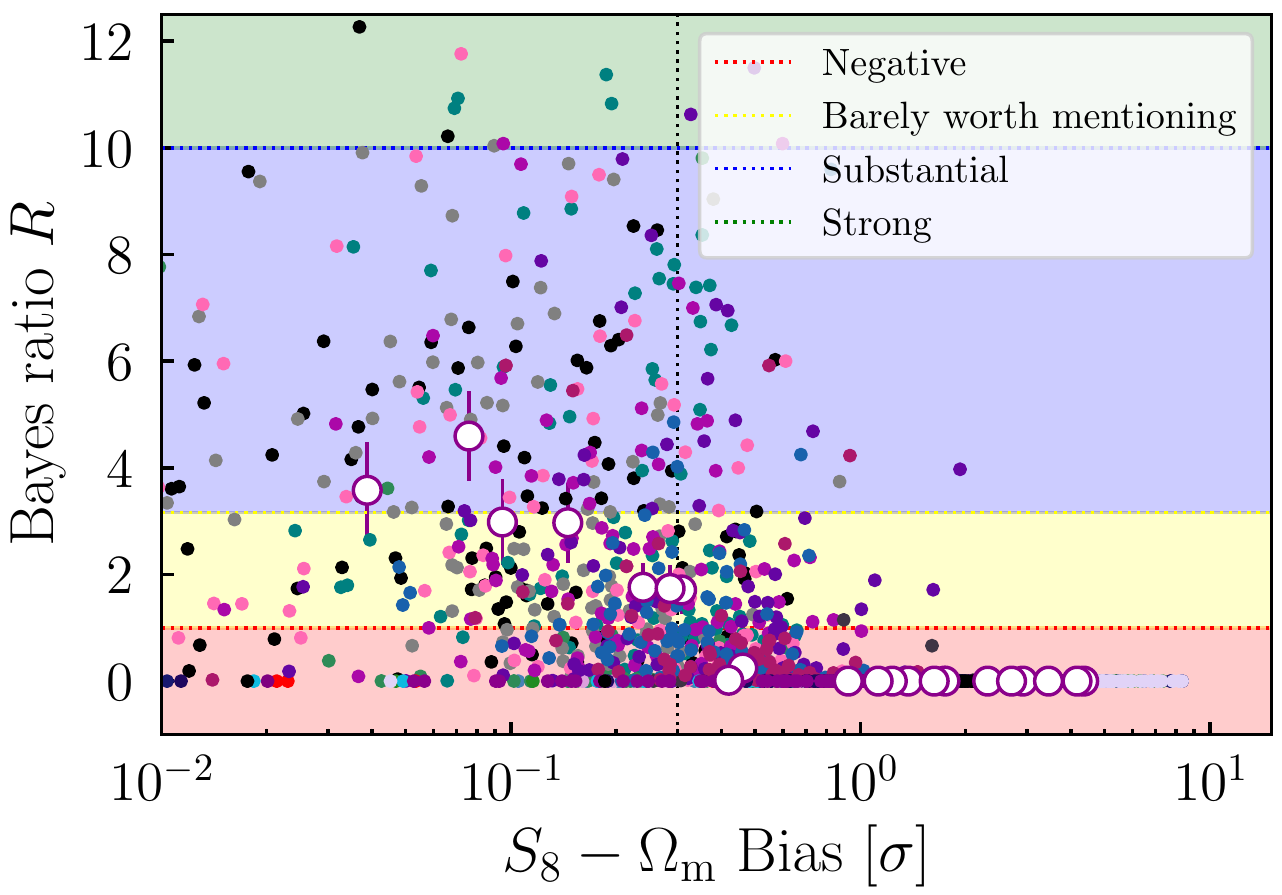}
\includegraphics[width=\columnwidth]{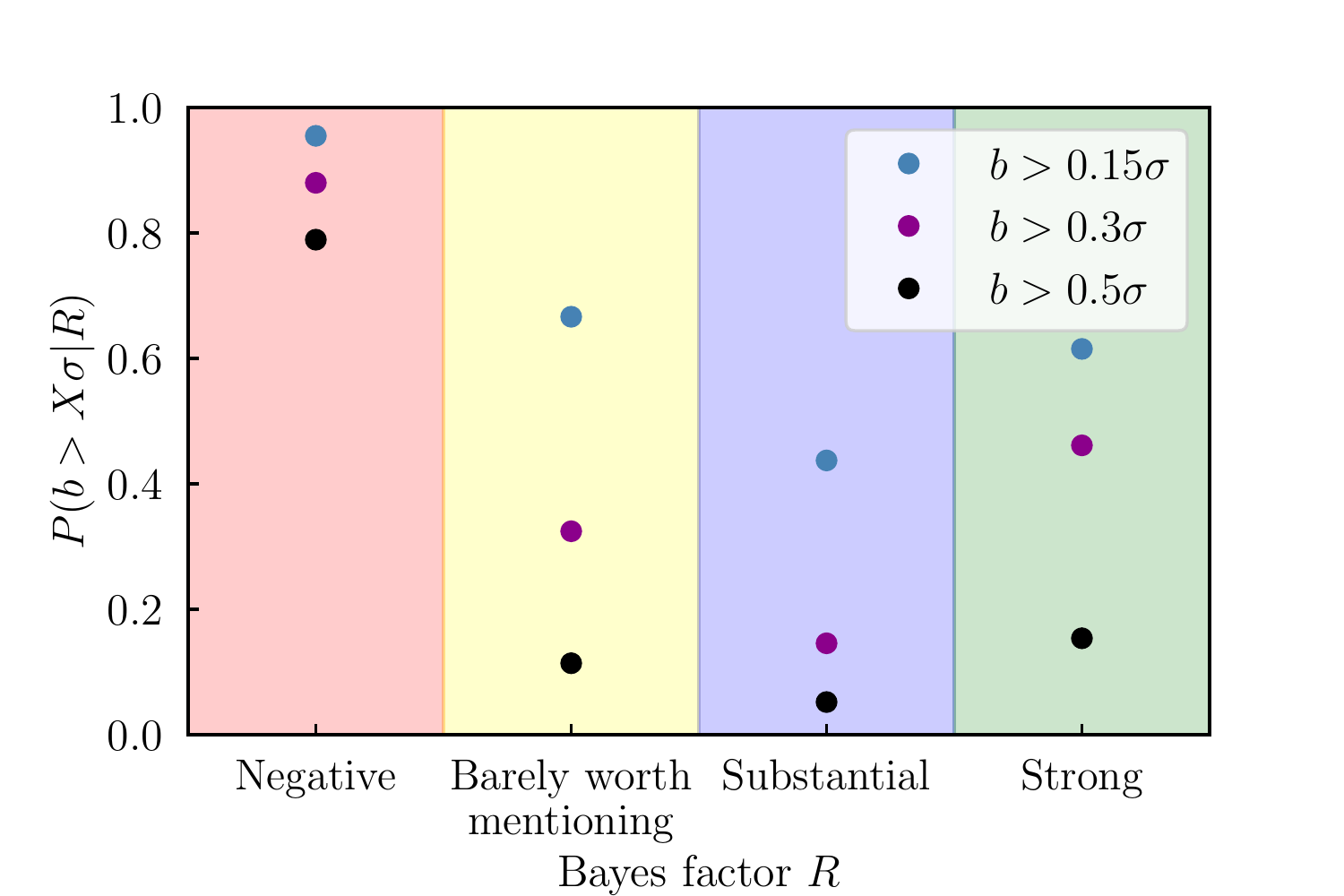}
\caption{\textbf{Top:} The same 21 samples as in Figure \ref{fig:bias_results}, but now showing the Bayes ratio $R=\mathcal{Z}_{\rm NLA}/\mathcal{Z}_{\rm TATT}$ rather than $\Delta \chi^2_{(\textrm{df})}$. As before, the open points show the bias/evidence ratios estimated by running \blockfont{PolyChord} on noiseless data vectors. The points represent the scatter due to noise (50 noise realisations for each IA scenario; see Section \ref{sec:data:noise} for details). The vertical line shows the $0.3\sigma$ bound, and the shaded colours show how the different ranges of $R$ are interpreted according to the Jeffreys scale.
\textbf{Bottom:}The same as Figure \ref{fig:bias_prob_chi2_threshold}, but showing the Bayesian factor $R$ (defined as the ratio of Bayesian evidence values obtained from running NLA and TATT on the same data). The coloured bands represent categories on the Jeffreys scale, and $P$ is the probability of more than $X\sigma$ cosmological bias in the NLA model, given an observed Bayes factor in each category.
\label{fig:bayes_bias_results}
}
\end{figure}

In Figure~\ref{fig:bayes_bias_results} we show the same 21 IA scenarios as in Figures~\ref{fig:bias_results} and~\ref{fig:bias_prob_chi2_threshold}, but now using the Bayes factor as our model comparison statistic. In the top panel, the open points again show the noiseless case, with evidence values computed using \blockfont{PolyChord}. As we can see, there is a weak correlation, with low bias scenarios tending to give somewhere between ``substantial" and ``barely worth mentioning" on the Jeffreys scale (the coloured bands). Interestingly, in none of our IA scenarios, not even in the regime that is functionally unbiased, do we see ``strong" evidence for NLA. 

The scattered points in the top panel of Figure~\ref{fig:bayes_bias_results} show the impact of noise, as estimated using importance sampling. Each colour represents an IA sample, with different points representing different noise realisations. Clearly $R$ is significantly more sensitive to noise than $\Delta \chi^2_{(\textrm{df})}$, as we can see by comparing Figures~\ref{fig:bayes_bias_results} and~\ref{fig:bias_prob_chi2_threshold}. 
We observe essentially two scenarios -- when the bias in the $S_8 - \omegam$ plane is greater than $\sim 1\sigma$, the Bayes Ratio can tells us that NLA is disfavoured relatively reliably. On the other hand, when the bias in the $S_8 - \Omega_m$ plane is smaller than $1\sigma$, there is a considerable amount of scatter.

The bottom panel in Figure~\ref{fig:bayes_bias_results} shows the bias probability, conditioned on the Bayes ratio category. That is, given the data return $R$ in a particular class on the Jeffreys scale, $P$ is the probability that NLA is biased by more than $X\sigma$. There is clearly at least some information here; if the Bayes factor actively favours TATT ($R<1$), there is a high probability that NLA will be biased, even factoring in noise. Values within the ``barely worth mentioning" category could essentially go either way. In the ``substantial" category things look better, but even here there is $\sim10-15\%$ chance of biases more than $0.3\sigma$, and almost $50\%$ chance that NLA is biased by more than $0.15\sigma$ in the $S_8-\omegam$ plane. 

We can perhaps understand the relative noisiness in $R$ by considering Eq.~\eqref{eq:approx_evidence}. Assuming a Gaussian likelihood, the Bayes factor scales as $e^{\Delta \chi^2}$; any small perturbation in $\chi^2$ due to sampling noise will thus be magnified exponentially.  
We cannot say from this whether this is an inherent issue with the Bayes ratio, or only when estimated using our method of importance sampling. In the absence of an alternative fast method to estimate $R$ for many noise realisations, however, we recommend $\Delta \chi^2_{(\textrm{df})}$ as a more robust metric to use with our method. 

\section{Sampler Comparison}\label{app:multinest_polychord}

\begin{figure}
\includegraphics[width=\columnwidth]{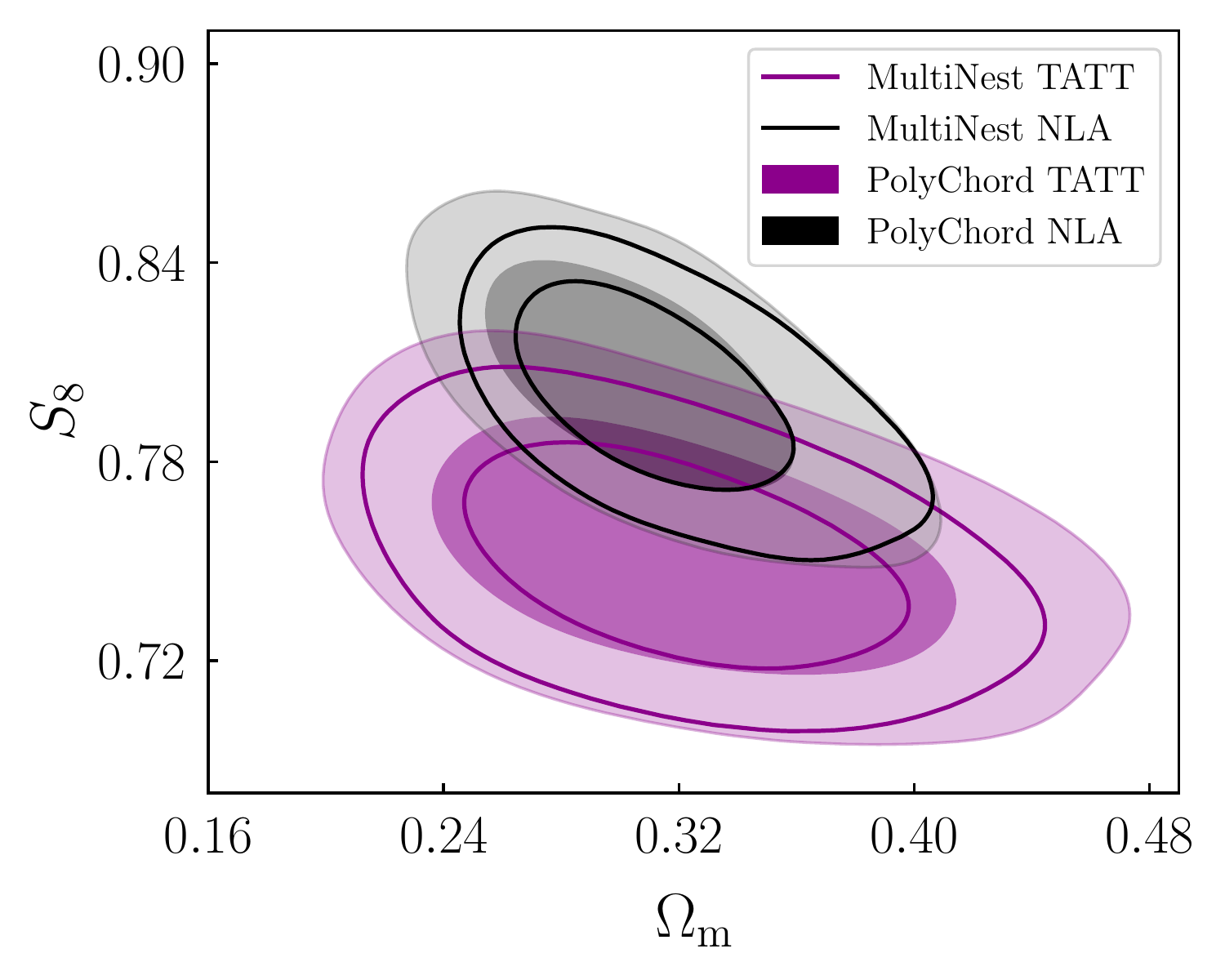}
\caption{Marginalised posteriors from a single noisy data vector, with a given input IA scenario. The shaded purple and black contours show the results of fitting that data assuming TATT and NLA respectively, using the \blockfont{PolyChord} nested sampling code. The unfilled contours are the results of the same analyses, but using the faster, but less accurate, \blockfont{MultiNest} algorithm.
}
\label{fig:s8om_samplers}
\end{figure}

In this appendix we present a brief comparison of two commonly used nested sampling codes: \blockfont{PolyChord} and \blockfont{MultiNest}. Although a similar (albeit more extensive) exercise is discussed in \citet*{y3-samplers}, their analysis choices differ significantly from ours, and so it is worth revisiting the question. To this end, we re-analyse our 21 noise $0$ IA data vectors using \blockfont{MultiNest} (500 live points, efficiency$=0.3$ , tolerance$=0.01$). The results are then compared with our fiducial \blockfont{PolyChord} run (500 live points, num\_repeats$=30$, tolerance$=0.01$).
We find:
\begin{itemize}
    \item The two samplers give consistent results for point estimates. That is, both can reliably locate the posterior mean, and the sampling around the peak is comparable, giving a similar level of noise in the best fit. As a result, the $\chi^2$ difference between NLA and TATT analyses is relatively insensitive to the choice of sampler. 
    \item \blockfont{MultiNest} is seen to underestimate the width of the $1\sigma$ posteriors on cosmological parameters significantly. This is true in both models; combined with the previous point, it leads to a systematic overestimation of the $S_8-\omegam$ bias for any given IA scenario. This can be seen in Figure \ref{fig:s8om_samplers}, which shows the posteriors as estimated by both samplers for a particular IA scenario.
    \item The Bayesian evidence estimates from \blockfont{MultiNest} are low compared with \blockfont{PolyChord}, as was shown in  \citet*{y3-samplers}. Although this is true for both models, the overall result is to increase $R$ (i.e., push the Bayes ratio towards favouring NLA more strongly).
\end{itemize}

\noindent
Given the observations listed above, we chose to use \blockfont{PolyChord} as our fiducial sampler, despite the runtime advantage of \blockfont{MultiNest}.

\section{Computational Resources}\label{app:computational_resources}

Here we describe the computational resources used for this paper. The aim is to provide an estimate of the computing power required to apply the method described in Section \ref{sec:method:recomendation} to perform model selection. The exact amount of time/resources will naturally vary depending on the details of the analysis pipeline. In our particular case:

\begin{itemize}
    \item Generate $\sim$ 25 IA samples: less than 1 minute in 1 core.
    \item Generate $\sim$ 25 IA datavectors: less than 1 minute per datavector in 1 core.
    \item Run 25 NLA chains using \blockfont{PolyChord}: around 22h per chain in 128 cores.
    \item Run 25 TATT chains using \blockfont{PolyChord}: around 28h per chain on 128 cores.
    \item Generate 50 noise realisations, to be added to each one of the 25 data vectors: less  than 1 minute in 1 core.
    \item Generate importance sampling weights and the $\chi^2$ pool for 25$\times$50 noisy data vectors: around 6-12h on 128 cores.
    \item Apply the weights to evaluate the NLA and TATT posteriors and compute the bias and best fit in each case: around 6-12h on 128 cores.
\end{itemize}

\end{document}